\documentclass[sigconf,10pt]{acmart}

\usepackage[ruled,vlined]{algorithm2e}
\usepackage{subcaption}
\usepackage[noend]{algpseudocode}
\usepackage{multirow}
\usepackage{cleveref}
\usepackage{siunitx}
\usepackage{enumitem}

\newcommand{\tool}{{Hera}}
\newcommand{\mir}{{MIR }}

\newcommand{\ccp}{{congestion control protocols\space}}
\AtBeginDocument{%
  }

\renewcommand\footnotetextcopyrightpermission[1]{} % removes footnote with conference info
\acmConference{}{}{}

\begin{document}

\title{Towards Next Generation Immersive Applications in 5G Environments}

% \affiliation{%

% \affiliation{%

\author{Rohail Asim}
% \email{rohail.asim@nyu.edu}
\affiliation{
    \institution{New York University}
    % \country{USA}
}

\author{Ankit Bhardwaj}
% \email{bhardwaj.ankit@nyu.edu}
\affiliation{
    \institution{New York University}
    % \country{USA}
}

\author{Lakshmi Subramanian}
% \authornote{Note}
% \email{lakshmi@nyu.edu}
\affiliation{%
  \institution{New York University}
    % \country{USA}
}

\author{Yasir Zaki}
% \authornote{Note}
% \email{yasir.zaki@nyu.edu}
\affiliation{%
  \institution{New York University Abu Dhabi}
    % \country{UAE}
}

\renewcommand{\shortauthors}{}

\begin{abstract}
The Multi-user Immersive Reality (MIR) landscape is evolving rapidly, with applications spanning virtual collaboration, entertainment, and training. However, wireless network limitations create a critical bottleneck, struggling to meet the high-bandwidth and ultra-low latency demands essential for next-generation MIR experiences. This paper presents Hera, a modular framework for next-generation immersive applications, comprising a high-level streaming and synchronization layer for AR/VR systems and a low-level delay-based QoE-aware rate control protocol optimized for dynamic wireless environments. The Hera framework integrates application-aware streaming logic with a QoE-centric rate control core, enabling adaptive video quality, multi-user fairness, and low-latency communication across challenging 5G network conditions. We demonstrate that Hera outperforms existing state-of-the-art rate control algorithms by maintaining up to 66\% lower latencies with comparable throughput performance, higher visual quality with 50\% average bitrate improvements in our analysis, and improved fairness. By bridging the gap between application-level responsiveness and network-level adaptability, Hera lays the foundation for more scalable, robust, and high-fidelity multi-user immersive experiences.

\end{abstract}

\maketitle

\section{Introduction}
\label{sec:introduction}

The landscape of Multi-user Immersive Reality (MIR) technology, encompassing both Virtual Reality (VR) and Augmented Reality (AR), is undergoing a transformative period driven by substantial investments from industry leaders like Meta and Apple \cite{apple_vision_pro_2024,apple_porsche_vision_2024,apple_health_vr_2024,meta_vr_investment_2024,meta_ai_ar_2024}. With the rapid growth of high-bandwidth \mir applications, the demand for ultra-fast and low-latency wireless communication has surged across both indoor and outdoor environments. In indoor settings, high-performance \mir applications often rely on Wi-Fi technologies such as IEEE 802.11ad (WiGig) and IEEE 802.11ay, which operate in the 60 GHz frequency band to enable multi-gigabit data rates. Similarly, outdoor \mir applications increasingly leverage 5G networks, particularly millimeter-wave (mmWave) 5G, to support real-time streaming and interaction. However, despite their potential for high-speed wireless connectivity, both Wi-Fi and 5G networks present fundamental challenges, including limited range, susceptibility to signal attenuation, and high bandwidth variability, all of which threaten the seamless performance of \mir applications.

% \subsection{Problem Setup}
% \label{sec:problem_setup}
% architecture figure
\begin{figure}
    \centering
    \includegraphics[width=0.95\linewidth]{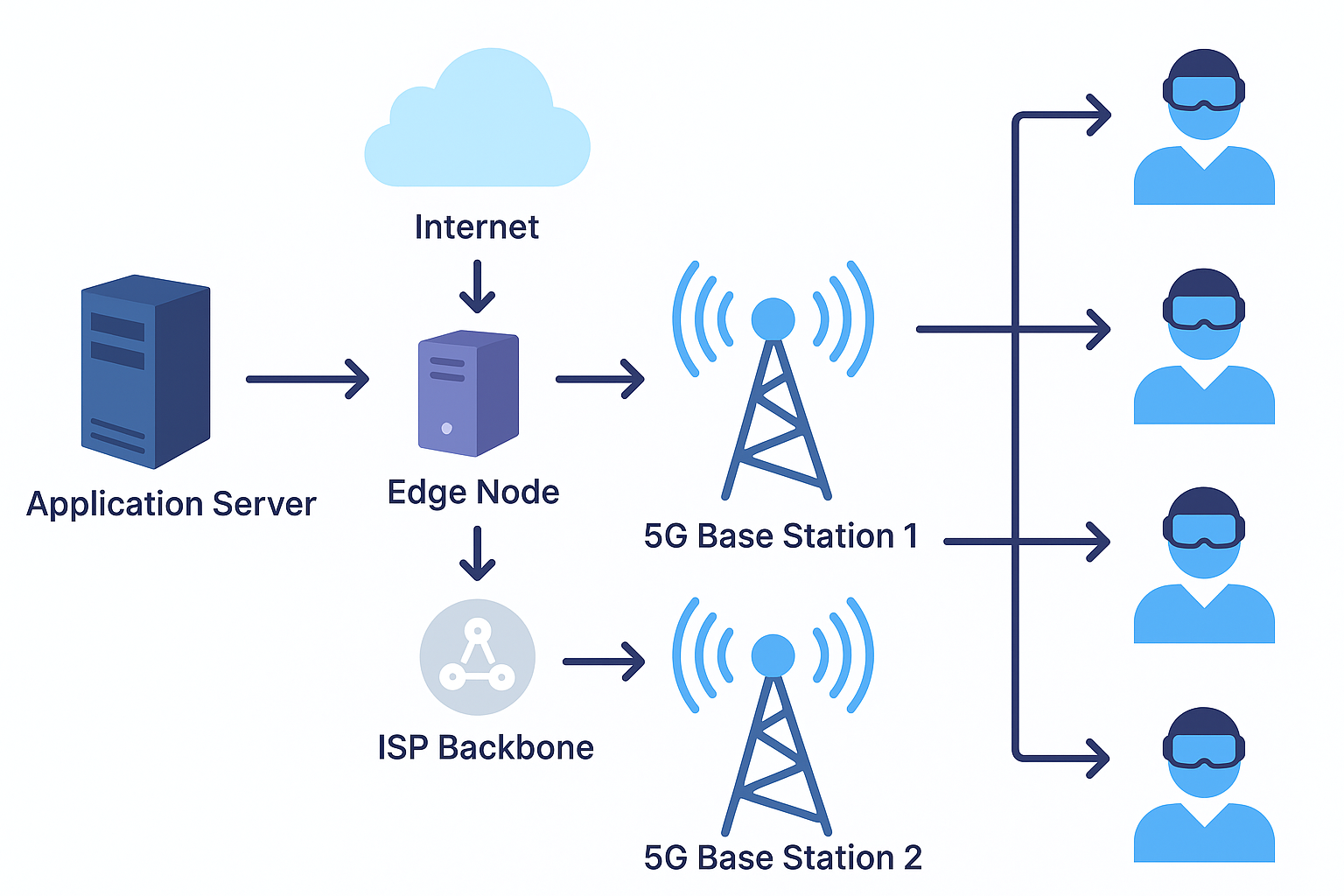}
    \caption{Real-world \mir application ecosystem. The end users can be indoors or outdoors and are connected to the application server by a 5G wireless channel.}
    \label{fig:arch}
\end{figure}

A critical yet often overlooked aspect of the infrastructure supporting multi-user immersive reality (MIR) applications illustrated in Figure~\ref{fig:arch} is whether existing systems, including transport protocols and application-level streaming architectures, can deliver the Quality of Experience (QoE) required for seamless interaction in dynamic wireless environments. Despite advances in mobile networking, the performance of \mir applications remains constrained not only by the limitations of traditional congestion control protocols but also by the lack of integration between network adaptation mechanisms and the application logic responsible for streaming, synchronization, and rendering. Addressing these issues requires a holistic framework that not only optimizes rate control at the transport level but also enables the application to dynamically adapt to changing network conditions to preserve the user experience.

% Current systems fail to adequately address the unique requirements of \mir experiences in modern wireless networks, where high variability in bandwidth, due to factors such as Wi-Fi interference indoors or mmWave instability outdoors, can compromise responsiveness and visual fidelity. These shortcomings directly impact key real-world QoE metrics that define user satisfaction in collaborative XR environments. Table~\ref{tab:qoe-metrics} summarizes several of these metrics, including startup delay, video resolution, interaction latency, and collaborative fluency. For example, high latency and throughput fluctuations lead to delayed interactions and lower-quality visuals, which in turn diminish immersion, introduce discomfort, and impair collaboration. Addressing these issues requires a holistic framework that not only optimizes rate control at the transport level but also enables the application to dynamically adapt to changing network conditions to preserve the user experience.

\begin{table}[h]
\centering
\begin{tabular}{p{2.5cm} p{5cm}}
\toprule
\textbf{QoE Metric} & \textbf{Description} \\
\midrule
\textbf{Startup delay} & The time users wait before the immersive content begins. Higher startup delay reduces perceived responsiveness at session start. \\
\textbf{Video bitrate / resolution level} & The visual clarity of the XR environment. Lower throughput forces the system to lower resolution or increase compression, reducing visual fidelity. \\
\textbf{Stall / buffering events} & Interruptions in the immersive experience where video or scene rendering pauses to rebuffer. Caused by throughput falling below content rate requirements. \\
\textbf{Interaction latency} & Delay between a user’s action (e.g., moving an object) and the visible response in the shared scene. Directly impacts perceived interactivity and collaboration smoothness. \\
\textbf{Collaborative fluency index} & The smoothness and synchronicity of shared actions among users, largely driven by latency. High latency leads to disjointed collaborative interactions. \\
\bottomrule
\end{tabular}
\caption{MIR QoE metrics and their user impact.\label{tab:qoe-metrics}}
\end{table}

\subsection{Challenges}
\label{sec:challenges}

\subsubsection{Application Requirements}
\label{sec:application_requirements}
Currently available \mir applications showcase some of the potential applications enabled by \mir but the Quality of Experience (QoE), in terms of key metrics summarized in Table~\ref{tab:qoe-metrics} such as video resolution, interaction latency, and collaborative fluency, for these applications is restricted by various network performance bottlenecks~\cite{firefly254406, Struye_2024, Perfecto_2020}. 
To circumvent these bottlenecks, many of these applications only support 2D experiences, which create a large ``screen'' in the immersive environment. Others that provide an immersive experience are forced to reduce the frames-per-second (FPS) for the application, which reduces how smooth the application feels. Alternatively, the application displays the content at a lower resolution, which reduces the graphical fidelity, or freezes momentarily, breaking the users' immersion \cite{10740019}. 
For a high-fidelity and comfortable \mir experience, a target frame rate (FPS) of at least 120 FPS is recommended, as lower frame rates can cause motion sickness due to the increased discrepancy between visual input and vestibular perception~\cite{10049694}. At the same time, 60 FPS and 90 FPS are often considered as a baseline and target for today's \mir systems~\cite{casasnovas2024experimental,10.1007/s00371-024-03501-4}.
In terms of resolution, a target video quality of at least 4K resolution per eye~\cite{Struye_2022} is required to mitigate the ``screen-door effect,'' a common VR artifact where inter-pixel gaps become visible and compromise the visual fidelity~\cite{cho201778}.

A high-quality, immersive, multi-user immersive reality experience featuring 360-degree 3D video at 120 frames per second (FPS) and 4K resolution per eye necessitates a downlink bandwidth of at least 100 Mbps per user to accommodate the high data rate associated with 360-degree video capture, stereoscopic 3D rendering, and real-time multi-user synchronization. Furthermore, end-to-end latency must remain below 20 ms to ensure a comfortable and responsive user experience, minimizing motion sickness and maximizing the sense of presence. This latency budget encompasses both network transmission delays and any processing overhead. Conventional congestion control algorithms, like BBR~\cite{cardwell2016bbr} and TCP~\cite{allman2009tcp}, are often insufficient to satisfy these stringent requirements. While UDP is commonly used for real-time communication due to its low overhead and minimal latency, many multi-user immersive applications require reliable delivery to ensure consistency across users, prevent visual artifacts, and maintain the integrity of complex shared scene states. TCP becomes essential in scenarios where packet loss or out-of-order delivery could disrupt synchronized interactions, collaborative object manipulation, or the seamless rendering of high-fidelity visual elements. BBR, while designed to maximize throughput, can exhibit performance degradation in lossy environments and may not consistently deliver the low latency necessary for VR.  Cubic~\cite{10.1145/1400097.1400105}, optimized for TCP flows, is similarly challenged by the real-time nature of VR streaming, as its congestion control mechanisms can introduce unacceptable delays, particularly in dynamic network conditions. The high bandwidth demands and strict latency constraints of high-fidelity VR necessitate the exploration and development of specialized network protocols and rate control algorithms tailored for real-time media streaming and interactive applications. 

\subsubsection{5G Channel Variability}
\label{5g_channel_variability}

% Variability in 5G environments
\begin{figure}[t]
    \centering
  \includegraphics[trim={0 0 17.93cm 0},clip,width=\linewidth]{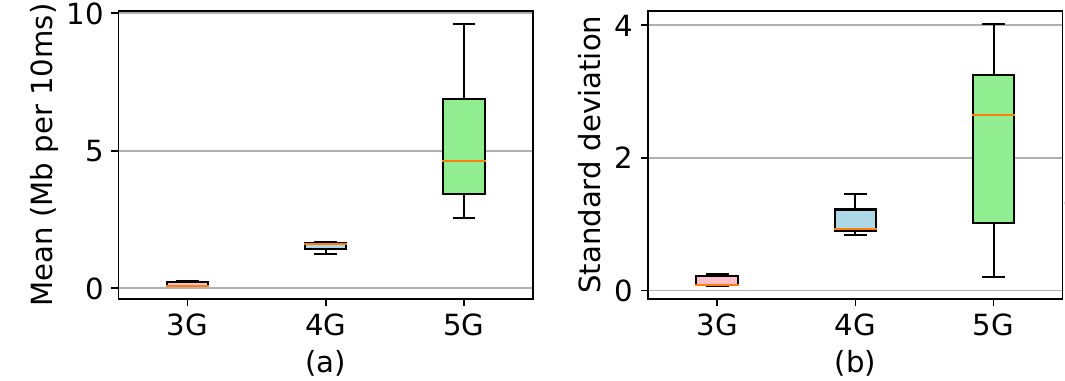}
  \caption{Cellular channels' variability analysis. The 5G channel provides higher mean bandwidth, but it  comes at the cost of higher bandwidth fluctuations. Data generated through experiments using real-world commercial networks.\label{fig:var_analysis}}
\end{figure}

Link capacity variation is a known fact in cellular networks. However, the nature of the 5G-New Radio (NR), with a sub-6 GHz spectrum and mmWave bands being vital elements, may cause link capacity fluctuations to be further amplified. Indeed, higher frequency bands suffer from larger propagation losses, effectively reducing the base stations’ coverage areas. For instance, due to weak diffraction ability, mmWave communications are sensitive to blockage by obstacles (e.g., humans, furniture, foliage). On the other hand, 5G networks are intended to support much higher mobility scenarios, supporting vehicular speeds up to 500 km/h. On the other hand, line of sight (LoS) and non-line of sight (NLoS) communications experience significantly different channel conditions and throughput. Although 3G and 4G also suffer from these channel fluctuations, these are far more accentuated in 5G links, as illustrated in Figure~\ref{fig:var_analysis} which provides perspective on the sheer difference in both bandwidth between 5G and non-5G networks and perhaps the bigger challenge, the high variability in 5G network environments that make it difficult for \ccp to adapt. Thus, despite the potential to enable high-throughput communications, the variability of the access channel capacity would result in a degraded TCP goodput and very low radio resource utilization. Existing \ccp are commonly designed as general-purpose protocols that consistently perform well in a wide variety of common scenarios. In MIR, however, applications have strict network requirements for optimal performance.

\subsection{Contributions}
\label{sec:contributions}

In order to overcome these constraints, we propose \tool, a modular framework for \mir applications, built around a novel QoE-aware rate control protocol optimized for next-generation wireless environments. The \tool\space framework integrates two layers: a high-level streaming and synchronization layer tailored for AR/VR applications and a low-level congestion control core that provides real-time network adaptation through delay-based window modulation and histogram-based RTT tracking. Together, these layers enable the system to dynamically adjust media quality, viewport streaming, and collaborative synchronization rates in response to network conditions, bridging transport-level adaptation and application-level quality of experience. Our work makes several key contributions. 

\begin{itemize}
    \item We design and implement an open-source framework that integrates transport-level rate control with application-level multi-user synchronization to support bandwidth-intensive AR/VR benchmarking. Our implementation connects a custom TCP-based kernel congestion control module with a WebXR-based multi-user application that synchronizes positional updates and interactions across both real headsets and synthetic clients. The framework is designed to expose network metrics to the application layer, providing a foundation for future extensions that could enable adaptive bitrate, resolution, field-of-view streaming, and dynamic synchronization rates.
    \item Our evaluation goes beyond traditional congestion control metrics by demonstrating how improvements in throughput, latency, and fairness translate into enhanced QoE for immersive applications. We show that \tool\space maintains low startup delay and stall frequency, sustains higher resolution levels, lowers interaction latency, and improves collaborative fluency compared to existing protocols such as BBR, Allegro, Vivace, and Cubic. These improvements directly impact the comfort, immersion, and usability of AR/VR systems.
    \item We demonstrate the performance improvements offered by the \tool\space rate control protocol using a system designed for adaptive video streaming to VR headsets, with a focus on measuring Quality of Experience (QoE). The system supports experiments with both real headsets and synthetic clients, enabling high scalability. It integrates a Linux server hosting DASH content via NGINX, VR and synthetic clients using \texttt{dash.js} for playback, and dynamic switching between rate control protocols. This setup allows controlled benchmarking of QoE under various load conditions and network environments. Our experiments show that \tool\space consistently outperforms other rate control protocols in delivering higher average video quality and smoother playback in multi-user immersive scenarios.

\end{itemize}

% table
\begin{table}[t]
\centering
\begin{tabular}{l S[table-format=2.2] S[table-format=1.3]}
\toprule
Protocol & {Average Quality} & {Fairness Index} \\
\midrule
BBR     & 34.5 & 0.989 \\
Cubic   & 12.2 & \textbf{0.999} \\
Allegro & 45.7 & 0.370 \\
Vivace  & 46.5 & 0.803 \\
\textbf{\tool}    & \textbf{91.2} & 0.965 \\
\bottomrule
\end{tabular}

\caption{Performance comparison of congestion control protocols. \tool\space achieves the highest quality while maintaining high fairness between competing flows in \mir applications deployed over a 5G network.\label{tab:protocol_comparison}}
\end{table}

\section{Related Work}
\label{sec:related}

\subsubsection*{AR/VR Applications in 5G Environments}
Many multi-user immersive reality (MIR) applications have been publicly released, but their operation within current network infrastructures often requires compromising Quality of Experience (QoE). For example, virtual social platforms like VRChat~\cite{vrchat} and Rec Room~\cite{recroom} enable users to interact in user-generated virtual worlds but face noticeable latency and graphical limitations. Collaborative design tools such as Spatial~\cite{spatial_io} and Arkio~\cite{arkio} support remote teams co-designing in shared 3D spaces, though they struggle with complex models and large user counts. Immersive training simulations for medical and industrial applications offer hands-on virtual practice but often require simplified visuals and limited interactivity to maintain real-time performance. Large-scale virtual events, such as concerts in Meta Horizon Worlds~\cite{metahorizon}, highlight both the promise of MIR applications and the challenges in delivering high-fidelity experiences to large audiences. Research addressing these challenges includes techniques like viewport prediction~\cite{xuviewport10.1145/3359989.3365413}, adaptive bitrate algorithms~\cite{ferranabr10.1145/3636534.3697322}, FOV streaming~\cite{hou9069299}, and resource-efficient multi-user AR frameworks like Spear~\cite{spear23.10.1145/3609395.3610596}. Despite these advances, scaling MIR applications to support seamless multi-user experiences in dynamic 5G environments remains an open problem.

\subsubsection*{Next-Generation Congestion Control Algorithms}
Numerous congestion control protocols have been proposed to overcome TCP's limitations in modern wireless environments. Machine learning-based solutions such as Remy~\cite{winstein2013tcp}, Indigo~\cite{yan2018pantheon}, Aurora~\cite{jay2019deep}, and Orca~\cite{abbasloo2020classic} aim to dynamically adjust sending rates under complex conditions, though they often struggle with generalization or stability. Delay-based protocols like Copa~\cite{arun2018copa}, Verus~\cite{zaki2015adaptive}, and Sprout~\cite{winstein2013stochastic} attempt to balance throughput and latency using end-to-end delay profiles or stochastic models, but typically require tight sender-receiver coordination. Real-time communication protocols such as Google Congestion Control (GCC)~\cite{gcc10.1145/2910017.2910605}, ScReAM~\cite{scream10.1145/2630088.2631976}, and NADA~\cite{nada6691448} are tailored for low-latency media but often exhibit limitations in bandwidth stability or infrastructure dependency. Meanwhile, transport innovations like QUIC~\cite{10.1145/3098822.3098842} and deep learning-driven systems like DeePCCI~\cite{10.1145/3341216.3342211} introduce further flexibility with encrypted header support and pluggable congestion control. Unlike these approaches, \tool\space integrates histogram-based RTT tracking with probabilistic window adjustment, targeting the specific demands of high-throughput, low-latency multi-user immersive applications in 5G environments.

\subsubsection*{QoE Metrics and Rate Control for AR/VR}
Understanding and optimizing QoE for AR/VR applications has led to the development of frameworks like VR-EXP~\cite{filho10.1145/3360286} and Perceive~\cite{filho10.1145/3204949.3204966}, which provide controlled environments and predictive models to assess streaming performance under variable network conditions. These platforms focus on metrics such as startup delay, stall frequency, and frame rate stability. Kulkarni et al.~\cite{kulkarni10.1145/3609395.3610599} evaluated Wi-Fi configurations for immersive video, highlighting key parameters that affect streaming QoE. While these works address aspects of AR/VR performance, they often stop short of directly linking transport-layer behavior to application-level QoE, particularly in the context of rate control mechanisms that can dynamically adjust to fluctuating bandwidth while maintaining fairness and responsiveness across users, a gap that \tool\space aims to fill.

\section{\tool\space Design}
\label{sec:\tool-design}

\subsection{Framework Design}
\label{sec:design}

The \tool\space framework is designed to provide end-to-end optimization and evaluation for multi-user immersive reality (MIR) applications operating over challenging wireless networks such as 5G.

\subsubsection*{High-Level AR/VR Application Layer}
Our high-level application layer is built around a flexible WebXR-based multi-user environment that supports real-time synchronization of user actions, such as positional updates and basic object interactions. This layer is designed to operate with both real VR headsets (e.g., Meta Quest, Pico, or Apple Vision Pro) and synthetic clients that emulate WebXR sessions. Real headsets allow us to validate performance and user experience under practical conditions, while synthetic clients enable large-scale, controlled benchmarking with customizable behavior profiles.

The system currently employs WebSockets for multi-user state synchronization, allowing position updates and interactions to be shared across clients with low latency. The application can be extended to incorporate WebRTC/WebTransport channels for bandwidth-intensive media delivery, adaptive bitrate management, and viewport-optimized streaming in future iterations. A key feature of the framework is its utility as a benchmarking tool. By modifying the sending rate, payload size, and update frequency of synthetic clients, the system can emulate a wide range of bandwidth-intensive applications, including high-fidelity video streaming, collaborative 3D design, and large-scale virtual events.

\begin{figure}
    \centering
    \includegraphics[width=0.5\linewidth]{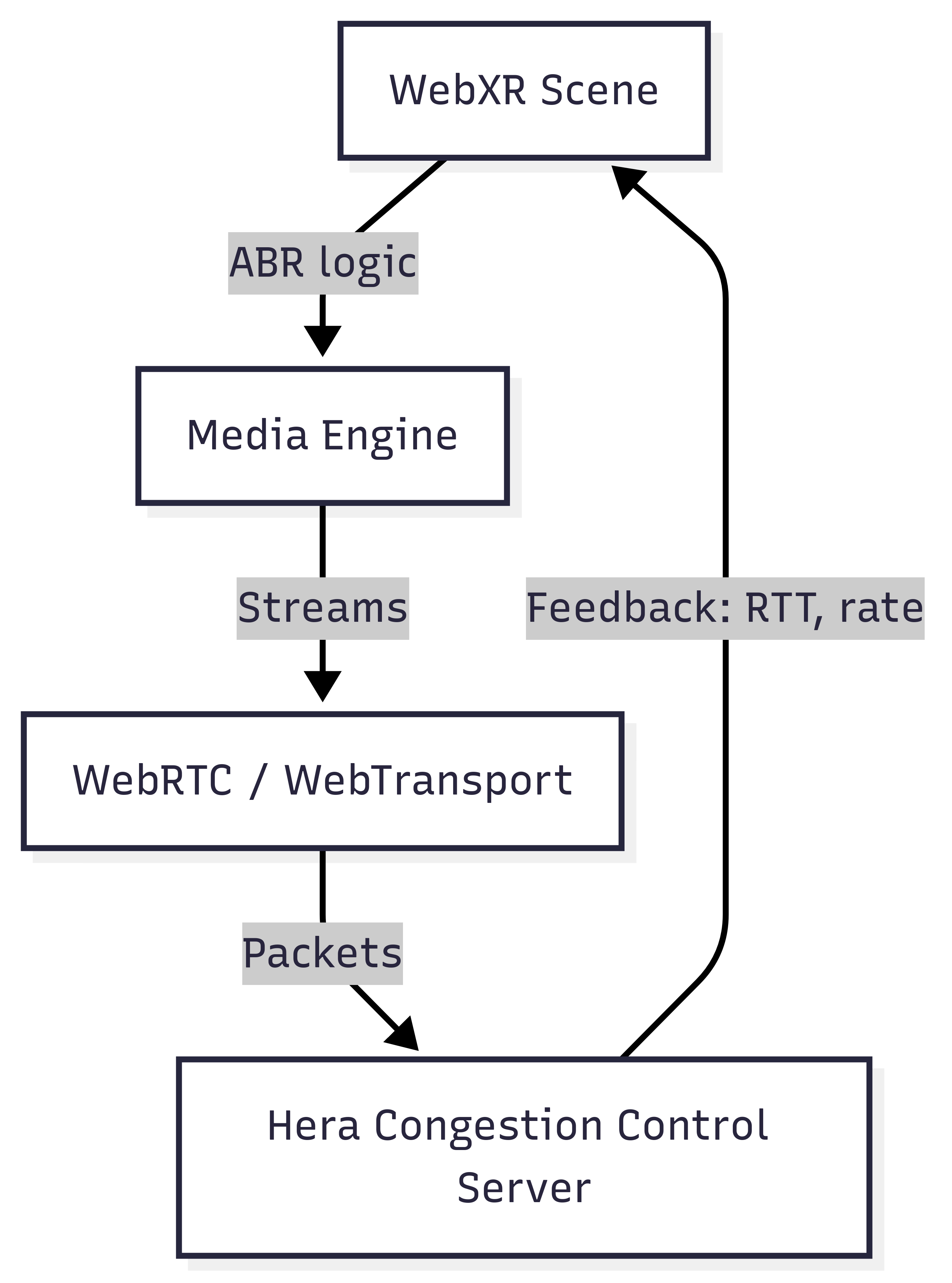}
    \caption{Framework architecture showing interactions between components.}
    \label{fig:framework_architecture}
\end{figure}

The \tool\space framework integrates seamlessly with our custom QoE-aware rate control protocol, implemented as a Linux kernel pluggable module. All application traffic, including multi-user state updates, synthetic client payloads, and configurable benchmarking streams, is transmitted over TCP flows managed by this module. This setup allows the selected rate control protocol to directly control sending rates and adapt to network conditions in real time. The framework is designed to expose network metrics and protocol feedback to higher layers, enabling future integration of congestion control feedback into adaptive application logic. Figure~\ref{fig:framework_architecture} provides an overview of the framework components and their interactions.

\subsection{QoE Aware Rate Control}

\subsubsection*{5G Network Setup}
The \tool\space rate control algorithm is designed for low-latency applications that operate in a 5G network environment where end hosts communicate with base stations over mmWave channels that exhibit high variability over short time scales~\cite{Rappaport2013Millimeter}. A critical requirement for congestion control in highly volatile network environments is that the protocol must adapt and converge towards the desired sending rate as quickly as possible, given that the environment state can change significantly within only a few RTTs. Given this requirement, the protocol design must consist of a simplistic algorithm that can make the required computations within microseconds while retaining high accuracy of predictions. Any organization deploying a 5G network will deploy several base stations within an area and operate a Radio Access Network (RAN) that can tightly manage all the radio allocations across base stations and end-hosts that connect to the network. We envision a setting where future 5G networks with edge compute infrastructure can support a broad array of low-latency applications with end-to-end latencies of less than $1-10$ms. In this regard, we assume that the server endpoint that connects to a low-latency application on a 5G mobile device is within the 5G network or within close network proximity to the 5G network. In summary, we assume a simple 5G network where an application from a mobile end-host connects with another endpoint (fixed or mobile) over a low-latency path where the primary network bottleneck is the highly variable wireless 5G network link. 

\begin{algorithm*}[!htb]
\caption{Rate Control Algorithm}\label{alg:hera}
\SetAlgoLined
\KwIn{Backlog \( \textrm{backlog} \leftarrow \textrm{FIFO queue of length N} \)}
\KwIn{Histogram \( \textrm{histogram} \leftarrow \textrm{Array[Number of buckets X]} \)}
\KwIn{Bucket size \( B \leftarrow \textrm{Size} \) (ms), \( \Delta_{max} \leftarrow 10 \)}
\For{each new RTT measurement}{
    $\textrm{backlog} \leftarrow \textrm{backlog} + [RTT]$ \Comment{Update FIFO} \\
    $\mu \leftarrow \frac{1}{N}\sum_{i=0}^{N} \textrm{backlog}[i]$ \Comment{Compute average} \\
    $b \leftarrow \lfloor \mu / B \rfloor$ \Comment{Determine bucket} \\
    $\textrm{histogram}[b] \leftarrow \textrm{histogram}[b] + 1$ \\
    $\alpha \leftarrow \frac{\sum_{i=0}^b \textrm{histogram}[i]}{\sum_{j=0}^X \textrm{histogram}[j]}$ \\
    
    \uIf{$b < \textrm{X}/2$}{
        $cwnd \leftarrow cwnd + (\alpha \times (X/2 - b) \times \Delta_{max})$ \\
    }\Else{
        $cwnd \leftarrow cwnd - (\alpha \times (b - X/2 - 1) \times \Delta_{max})$ \\
    }
    % $cwnd \leftarrow \max(minimum, \min(maximum, cwnd))$ \Comment{Clamp values} \\
}
\end{algorithm*}

\subsubsection{Rate Control Algorithm}
\label{sec:rate-control}

The rate control algorithm, outlined in Algorithm~\ref{alg:hera}, employs a delay-centric approach to maintain low network latency while preserving throughput stability. At its core, the algorithm dynamically adjusts the congestion window (cwnd) by analyzing real-time Round-Trip Time (RTT) distributions through two primary components: a sliding window backlog of length $N$ and a histogram-based delay classification system that converges towards the desired RTT, based on the number of buckets \textit{X}. For each incoming RTT measurement, the algorithm updates a fixed-length FIFO queue (backlog) of $N$ recent RTT values, computes their moving average, and maps this average to a histogram bucket representing discrete latency ranges (e.g., 15 ms intervals). The histogram tracks the frequency distribution of these buckets over time, enabling the calculation of $\alpha$—a normalized metric reflecting the cumulative probability of observing RTTs in the current or lower-latency buckets. Based on this probability and the bucket index, \tool\space modulates the congestion window asymmetrically: when operating in low-latency regimes (buckets below a predefined threshold), it increases the window proportionally to both $\alpha$ and the distance from the threshold, fostering aggressive utilization of available bandwidth. Conversely, in high-latency states, it reduces the window based on the severity of the observed delay, effectively curbing queue buildup. The algorithm incorporates safeguards to clamp the congestion window within empirically validated bounds, preventing extreme oscillations. In simple terms, the maximum change in sending rate occurs when a large number of recently observed RTTs are significantly different from previous observations, and these RTTs fall in the furthest buckets from the center. This dual mechanism—probabilistic delay classification coupled with gradient-based window adaptation enables \tool\space to preemptively mitigate congestion before packet loss occurs, making it particularly effective for latency-sensitive applications such as real-time video streaming, cloud gaming, and \mir applications where stable, low-delay communication is critical.

\subsubsection{\tool\space rate control parameters}
\label{sec:\tool-parameters}

\begin{figure}
    \centering
    \includegraphics[width=\linewidth]{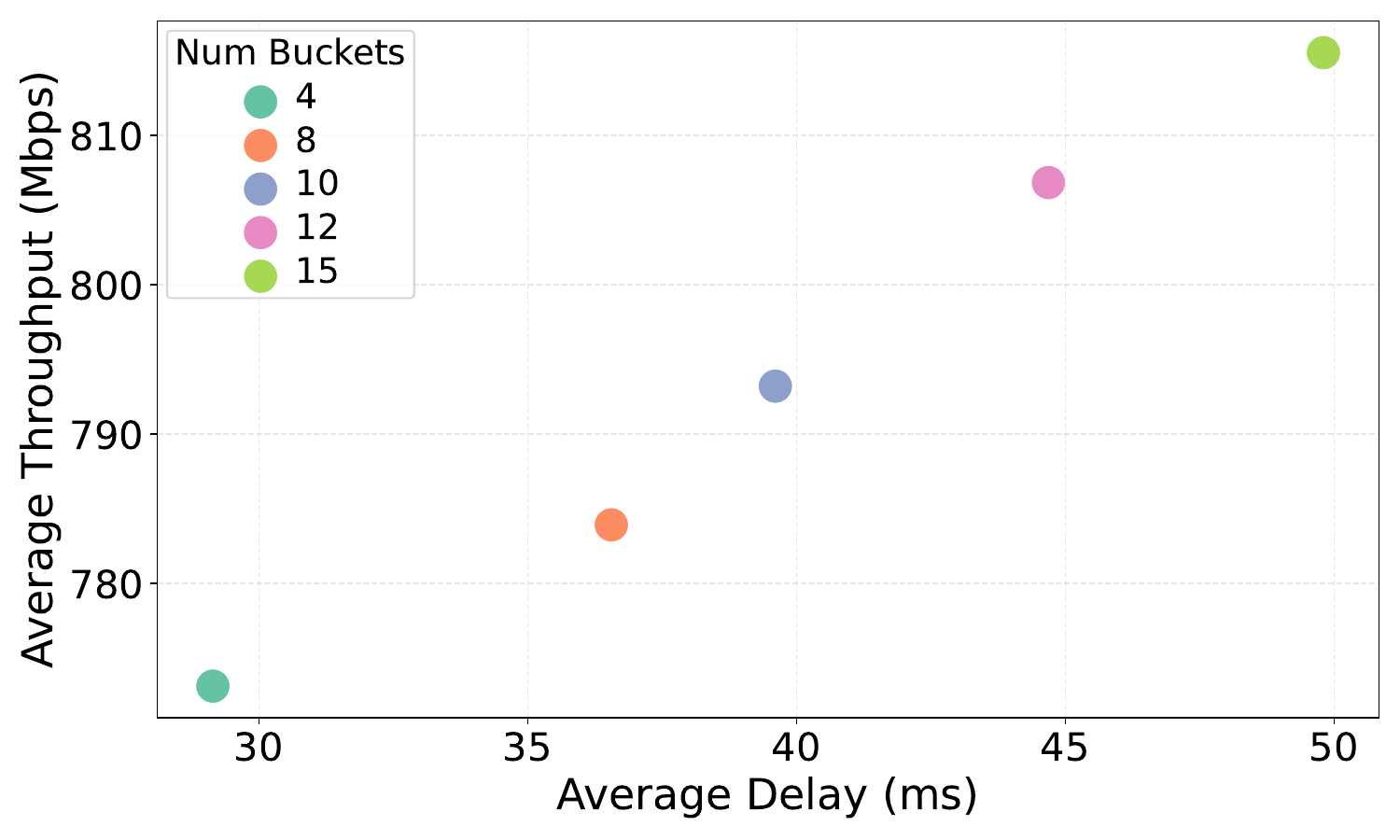}
    \caption{Impact of varying the number of buckets on protocol performance, measured in average throughput and delay. Results show that increasing the bucket count raises throughput but also increases delay.}
    \label{fig:numbuckets}
\end{figure}

\tool\space requires four control parameters that drive its performance, i.e., Bucket Size, the Max Delta value, Histogram Limit, and the \textit{backlog length}. These parameters can be finely tuned concerning the channel environment for enhanced performance gains.

\subsubsection*{Number of Buckets} 
The number of Buckets in the histogram determines the target RTT for the protocol. For a stable network environment, a low number of buckets is required, and for highly fluctuating network environments, a high number of buckets would be more suitable. For the general case, this value needs to be dynamically modified based on the network environment. For the VR streaming over 5G environments case, we can calculate a reasonable bucket size based on the RTT requirements of the VR applications for a smooth VR streaming experience. We can see the effects of changing the number of buckets in figure~\ref{fig:numbuckets}. As expected, the algorithm modifies the congestion window to ensure that the RTT converges towards the target RTT at the center of the distribution. For applications that prioritize higher throughputs with tolerance for higher RTTs, this parameter can be tuned to a higher value accordingly.  

\subsubsection*{Max Delta} 
The maximum delta determines the maximum value by which the $cwnd$ is modified at every RTT. It is used in combination with the current state and the alpha value, i.e., the probability of being in the current RTT or lower, to ensure that when the RTT is too high or too low, the cwnd is adjusted rapidly to converge towards the desired RTT. When the current state is close to the desired RTT, the protocol makes small adjustments to the cwnd. Increasing this value can cause the protocol to become unstable, as it may overreact to a signal and then change too rapidly in the opposite direction when attempting to recover and stay in this unstable loop.

\subsubsection*{Histogram Limit} 
The histogram limit is the maximum amount of data stored in the histogram, which is not only required for memory usage purposes but also to ensure that the global state of the network is not dependent on outdated data, which is not representative of the current network environment, as the state of the network environment can vary over time, especially for cellular networks.

\subsubsection*{Backlog length} 
The backlog stores a ``recent history'' of RTTs (Round-Trip Times). It stores the last few RTT measurements (e.g., the last 10 samples) in a sliding window. This helps smooth out temporary spikes or dips in delay. For example, if one RTT is unusually high (e.g., due to a random network hiccup), the backlog averages it with other recent RTTs to avoid overreacting. The backlog provides a short-term view of the network's current state, which is crucial for making quick, adaptive decisions. Increasing the backlog length would make the protocol slow to react to changes in the network environment, while decreasing it would make the protocol overreact to any observations.

\section{Evaluation Methodology}
\label{sec:methodology}

\subsection{VR Application}
In order to compare the \mir application performance of different congestion control protocols, we develop a system for streaming adaptive video to a VR headset and measuring Quality of Experience (QoE). The system is built around three core components: a Linux server hosting DASH content via NGINX, a VR client using dash.js for playback, and dynamic switching of congestion control protocols.

The server is configured to host DASH streams using NGINX with an RTMP module, which hosts video content as chunks (e.g., 4–10 seconds) along with a Media Presentation Description (MPD) file for adaptive bitrate streaming. Videos are pre-processed into multiple bitrate representations (0.045–4 Mbps) using FFmpeg and MP4Box to ensure DASH compatibility. To test different congestion control protocols, the Linux kernel's TCP stack is adjusted using sysctl commands (e.g., switching between CUBIC and BBR). Experiments involve streaming a 10-minute video (e.g., Big Buck Bunny) to the VR headset under controlled network conditions with five clients streaming the video simultaneously.

On the client side, five clients run the stream using a modified dash.js player. The player fetches the MPD file and adaptively selects video segments based on real-time network conditions. The dash.js player logs client-side Quality of Experience (QoE) metrics, including frame losses, bitrate, and video quality, to gauge application performance from the user's perspective. This setup allows us to systematically compare five different TCP congestion control protocols, including Hera, to assess how each protocol influences the performance of the VR application in terms of network efficiency and user experience. In order to emulate the variable latency, packet losses, and capacity limits that may be observed in real network environments, we use the Linux Traffic Control (tc) and NetEm tools. This approach provides valuable insights into the impact of different network protocols on a VR application, offering a practical foundation for optimizing VR experiences in varied network conditions.

\subsection{Network Trace Collection}
\label{sec:wigig-traces}

\begin{figure*}[!htb]
  \subfloat[City Drive\label{fig:trace_citydrive}]{\includegraphics[width=0.3\linewidth]{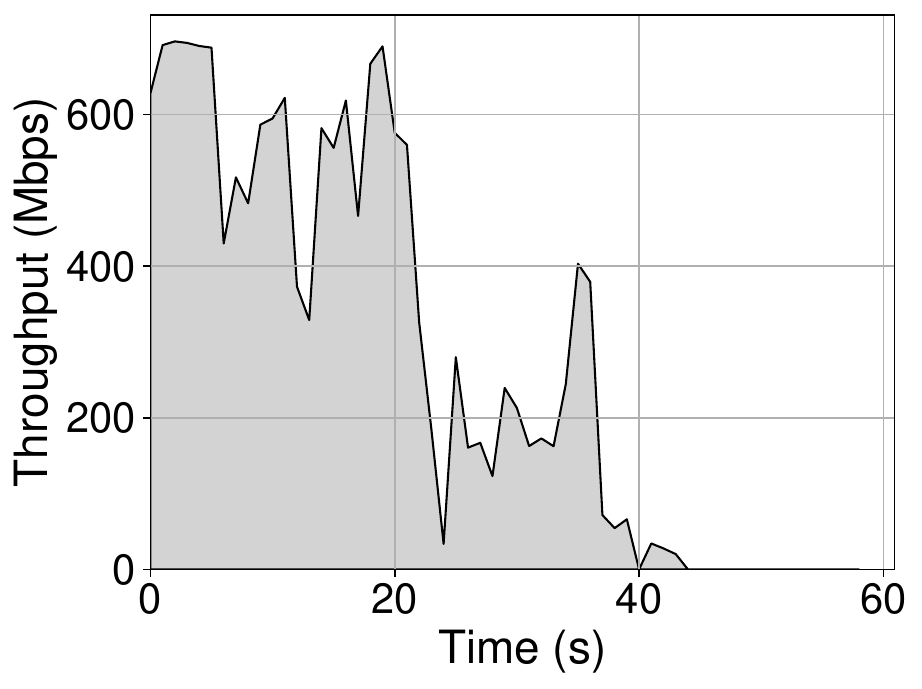}}\hfill
   \subfloat[Beach Stationary\label{fig:trace_beachstationary}]{\includegraphics[width=0.3\linewidth]{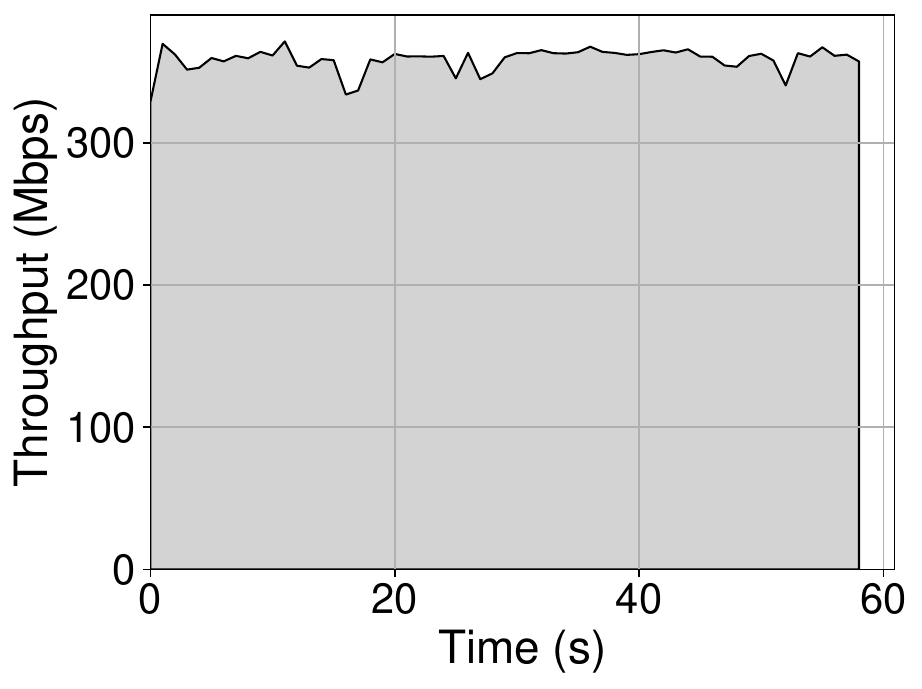}}\hfill
    \subfloat[RMa\label{fig:trace_ns3rma}]{\includegraphics[width=0.3\linewidth]{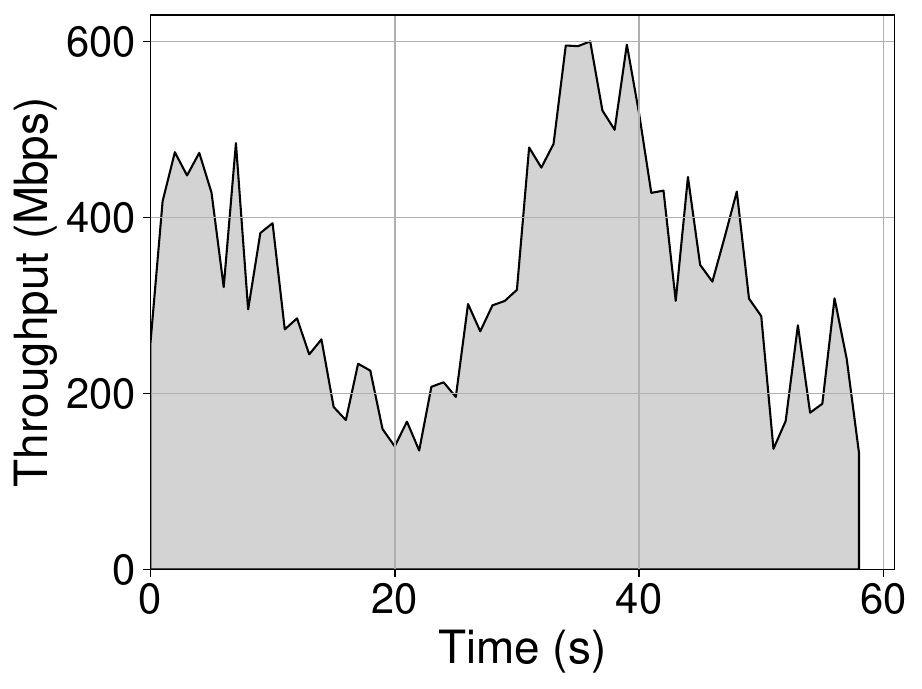}}\hfill
    \subfloat[UMa\label{fig:trace_ns3uma}]{\includegraphics[width=0.3\linewidth]{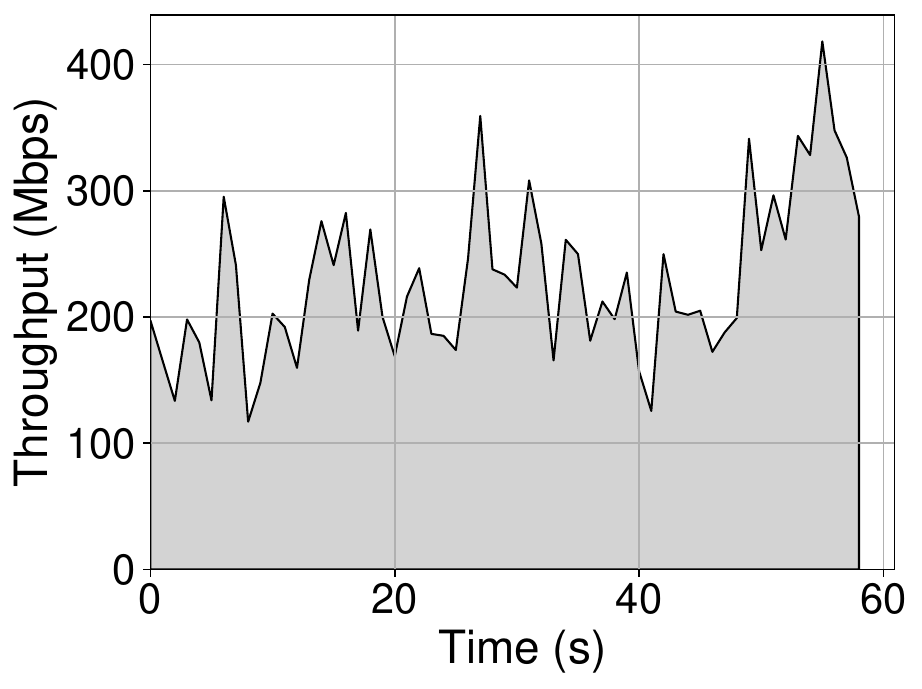}}\hfill
  \subfloat[Indoor Walking\label{fig:trace_wigigwalk}]{\includegraphics[width=0.3\linewidth]{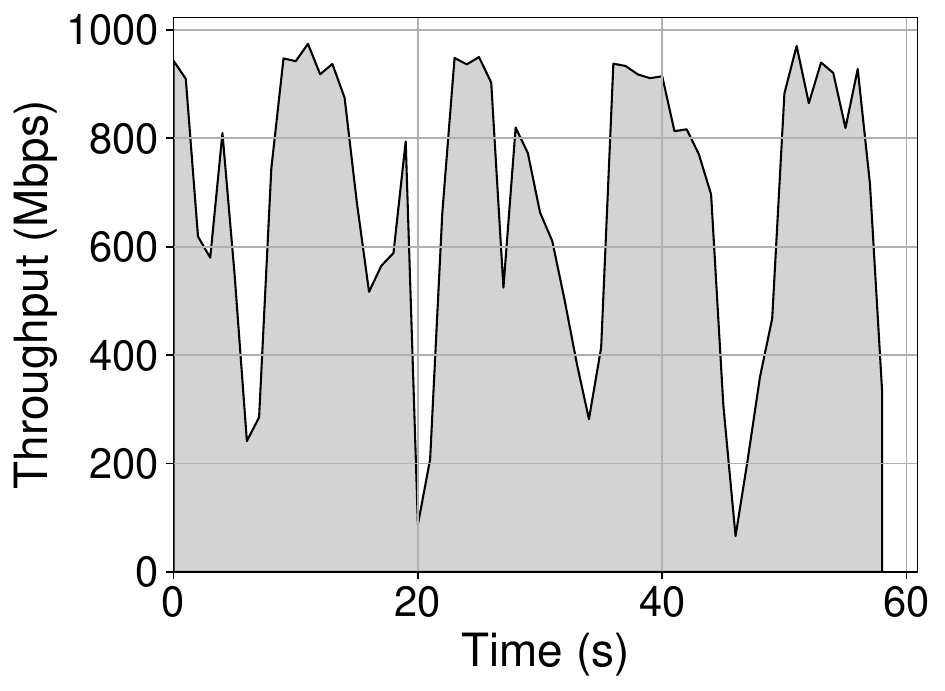}}\hfill
  \subfloat[Indoor Stationary\label{fig:trace_wigigstationary}]{\includegraphics[width=0.3\linewidth]{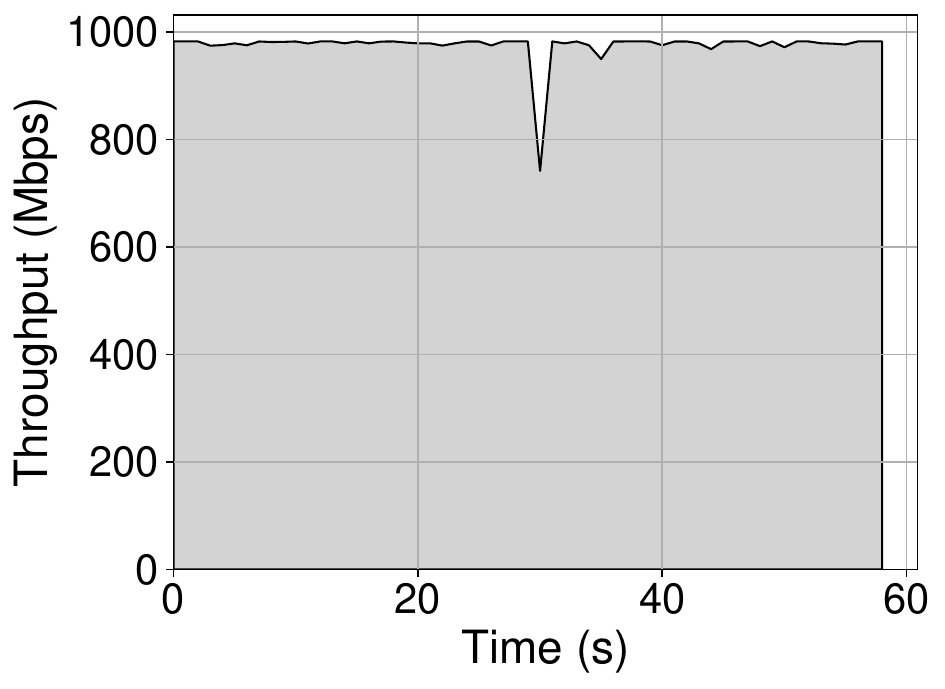}}\hfill

  \caption{5G network traces used for experiments were collected using a commercial 5G deployment, NS-3 mmWave simulations, and WiGig-based indoor setups, covering conditions MIR applications may encounter in real-world deployments. The traces include urban driving (City Drive, UMa), stationary scenarios (Beach Stationary, Indoor Stationary), rural mobility (RMa), and high-mobility indoor environments (Indoor Walking). \label{fig:channeltraces}}
\end{figure*}

In order to emulate a diverse set of realistic 5G network environments, we collect a set of channel traces that collectively cover a wide range of scenarios in terms of available bandwidth, mobility (driving, walking, and stationary), and variability. We employ different methodologies to generate this set of channel traces. Firstly, similar to prior work~\cite{winstein2013stochastic, zaki2015adaptive}, we use a real 5G connection from a commercial deployment in the wild, along with an Android application capable of recording the maximum available channel capacity by connecting to a remote server hosting dummy content specifically for this channel trace collection. These traces include urban driving and standing stationary at a beach, and the full channel traces are depicted in Figure~\ref{fig:trace_citydrive} and Figure~\ref{fig:trace_beachstationary}. Secondly, we use the mmWave module~\cite{Mezzavilla18} built atop the NS-3~\cite{ns3} tool developed by New York University wireless group (NYU Wireless) to generate channel traces emulating scenarios involving random user movement in an urban scenario, depicted in Figure~\ref{fig:trace_ns3uma}, and a rural environment, depicted in Figure~\ref{fig:trace_ns3rma} using a spatial channel model~\cite{zugno2020implementatio}. Finally, we use channel traces generated using a high-speed WiGig router in an indoor environment in scenarios with high user mobility, depicted in Figure~\ref{fig:trace_wigigwalk}, and in a stationary scenario, shown in Figure~\ref{fig:trace_wigigstationary}. Altogether, these channel traces encompass a broad range of realistic network scenarios that a user of a \mir application may encounter in the real world.

\subsection{Emulating the 5G traces}

We used the Mahimahi framework~\cite{netravali2015mahimahi} to emulate 5G network links based on the collected traces. 
% These trace files contain the number of slots available in any given millisecond to send a packet. 
Mahimahi's linkshell acts as a controlled router that queues packets and throws them at the desired rate as dictated by the trace file. The traces give us the ability to run several algorithms and compare the performance across scenarios in a unified and controlled manner without having to control for many of the external parameters that are usually faced by running things in the wild, i.e., unpredictable competing traffic from other users or random obstacles and user movements in the environment. A similar approach has been used by Verus~\cite{zaki2015adaptive} and Sprout~\cite{winstein2013stochastic} in the past.

We have mimicked the popular \emph{iperf} utility for our experiments, having separate sender and receiver programs. The receiver runs inside a Mahimahi linkshell, while the sender repeatedly transmits fixed-size blocks of 128 KiB to the receiver for 60 seconds. 
We chose $bucket\_size=15$, $backlog\_length=10$ and the $Max\_delta=10$, as our default parameters for \tool. We choose 4 prominent CC protocols for comparison in our experiments. These protocols include BBR, Allegro, Vivace, and Cubic.

\begin{figure*}[!htb]
  \subfloat[City Drive]{\includegraphics[width=0.3\linewidth]{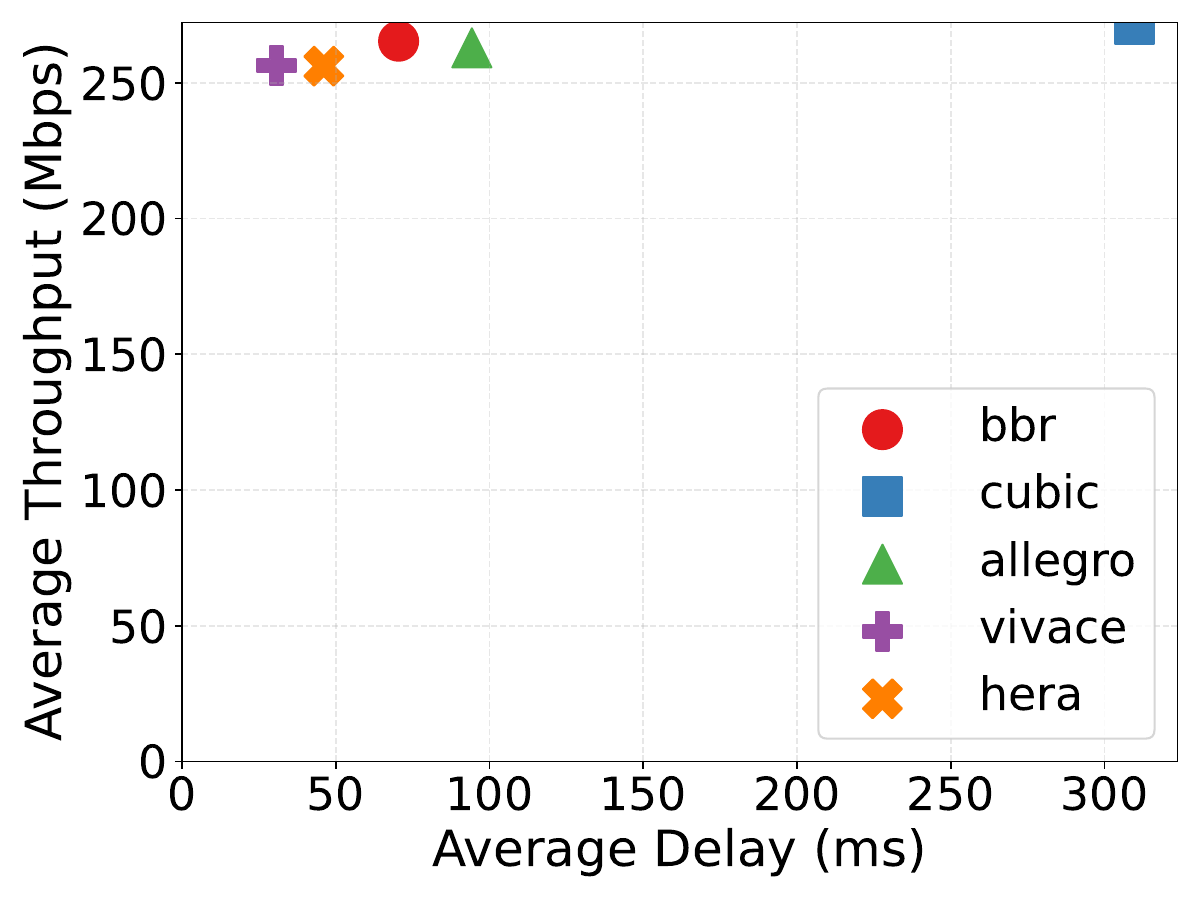}\label{fig:citydrive7}}\hfill
   \subfloat[Beach Stationary]{\includegraphics[width=0.3\linewidth]{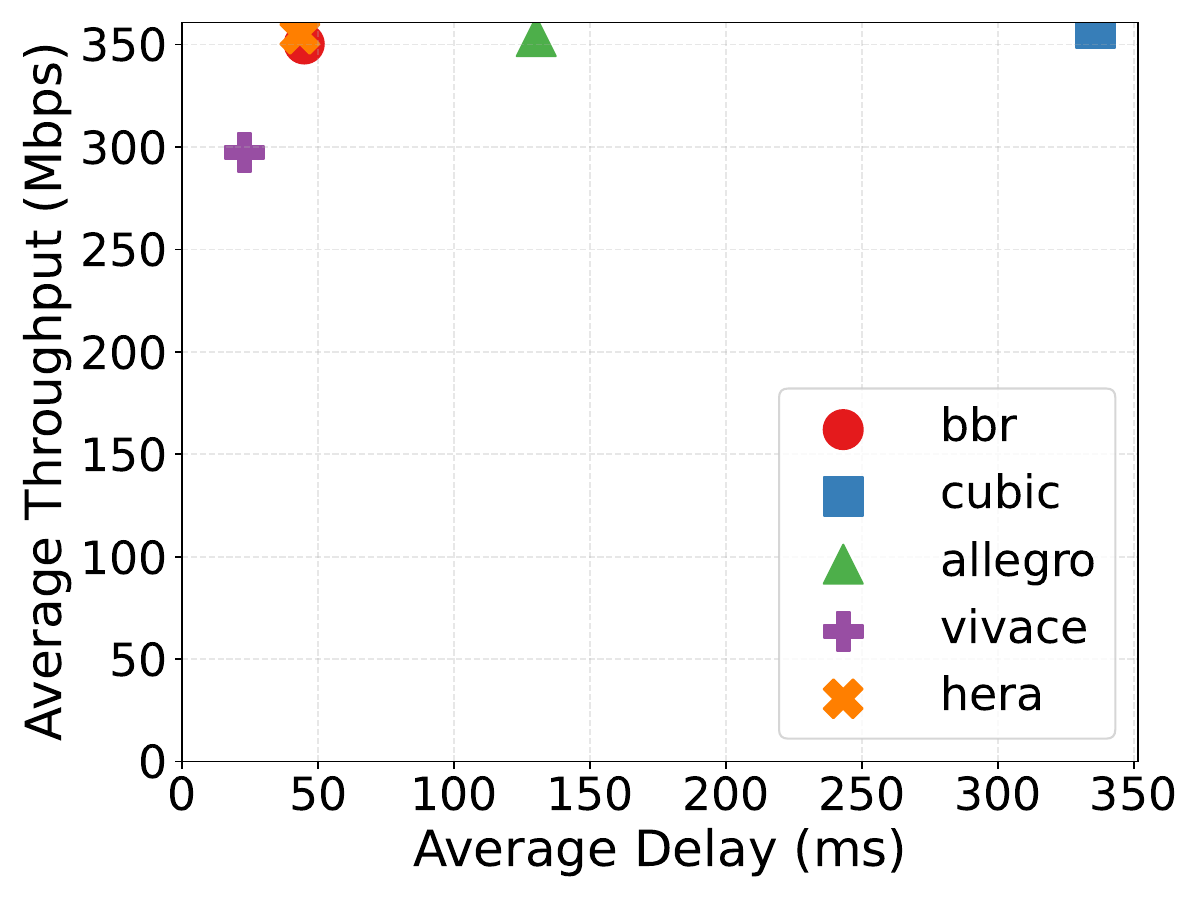}\label{fig:beach}}\hfill
    \subfloat[RMa]{\includegraphics[width=0.3\linewidth]{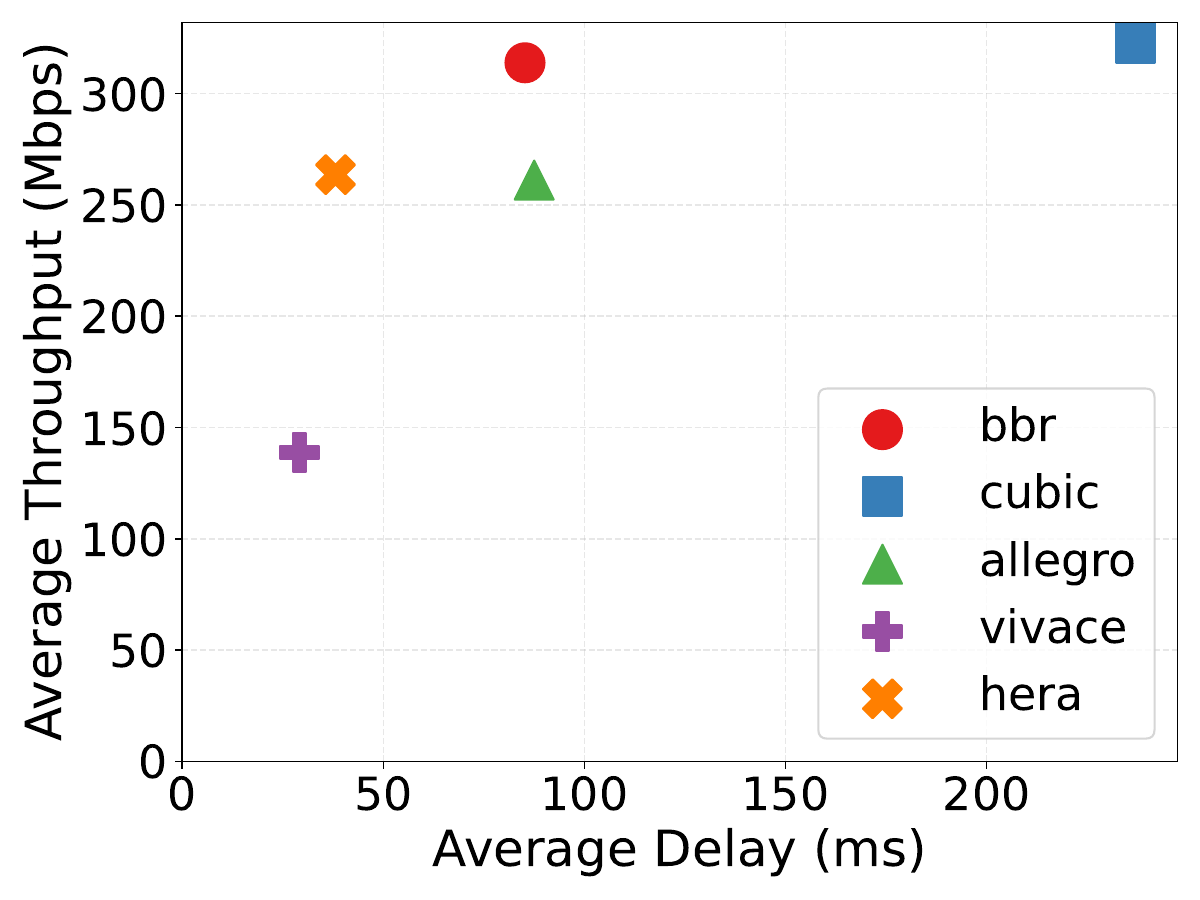}}\label{fig:rma}\hfill
    \subfloat[UMa]{\includegraphics[width=0.3\linewidth]{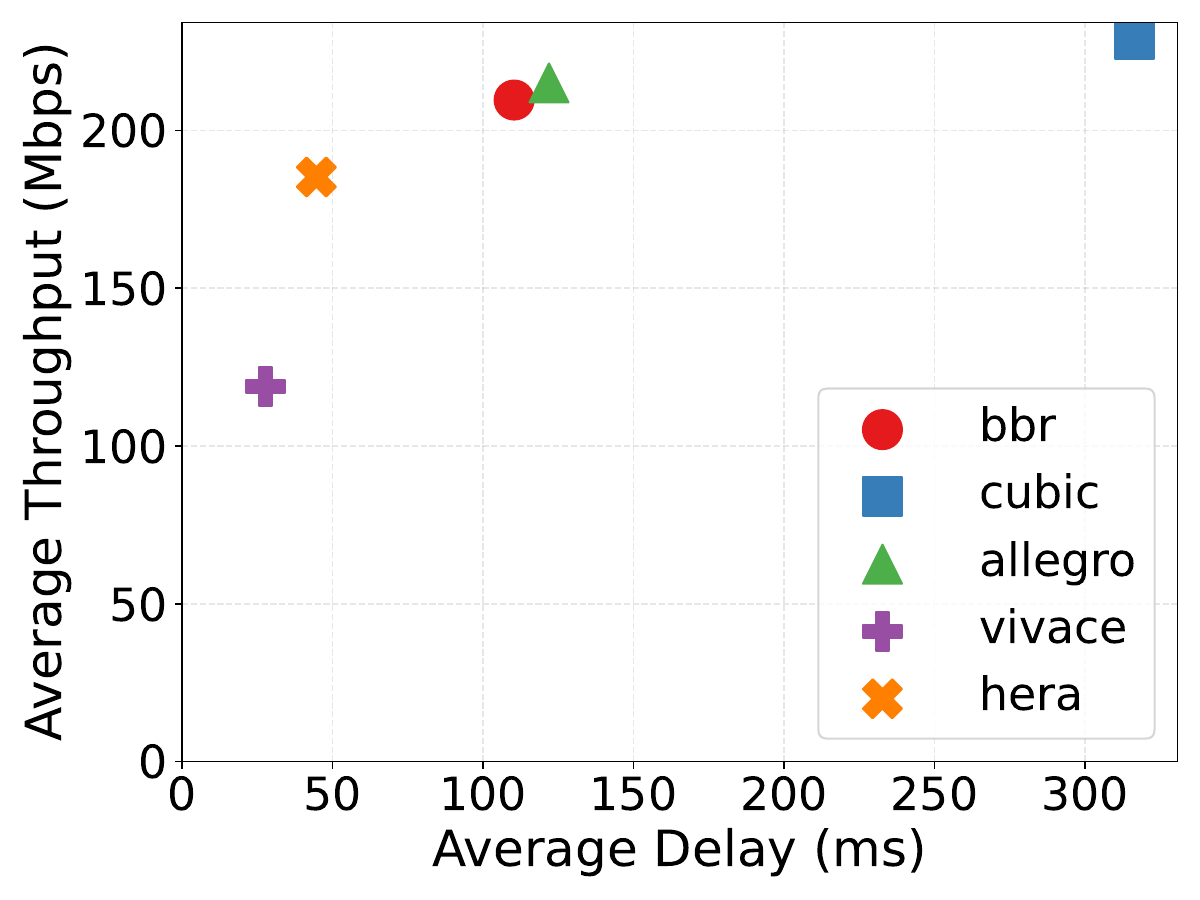}}\label{fig:uma}\hfill
  \subfloat[Walking]{\includegraphics[width=0.3\linewidth]{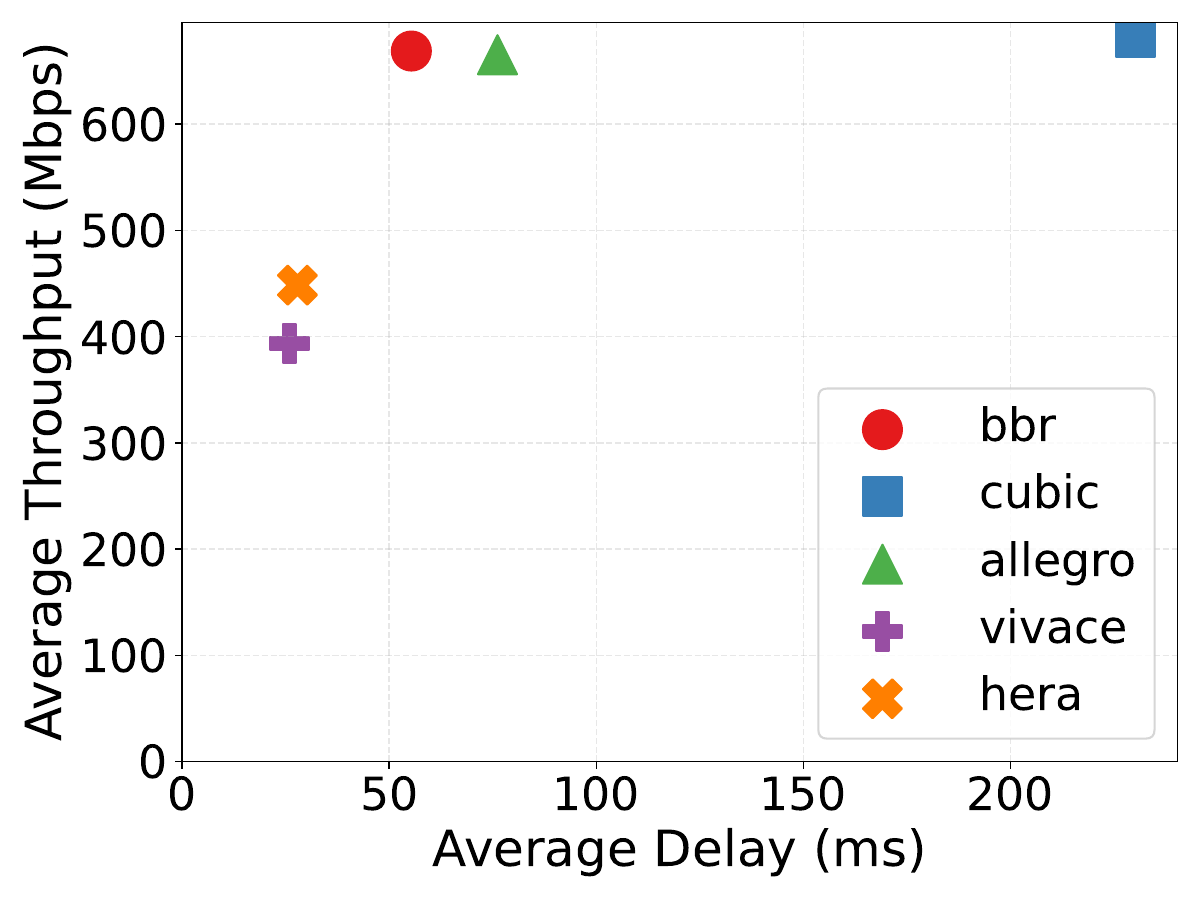}}\label{fig:walking}\hfill
  \subfloat[Stationary]{\includegraphics[width=0.3\linewidth]{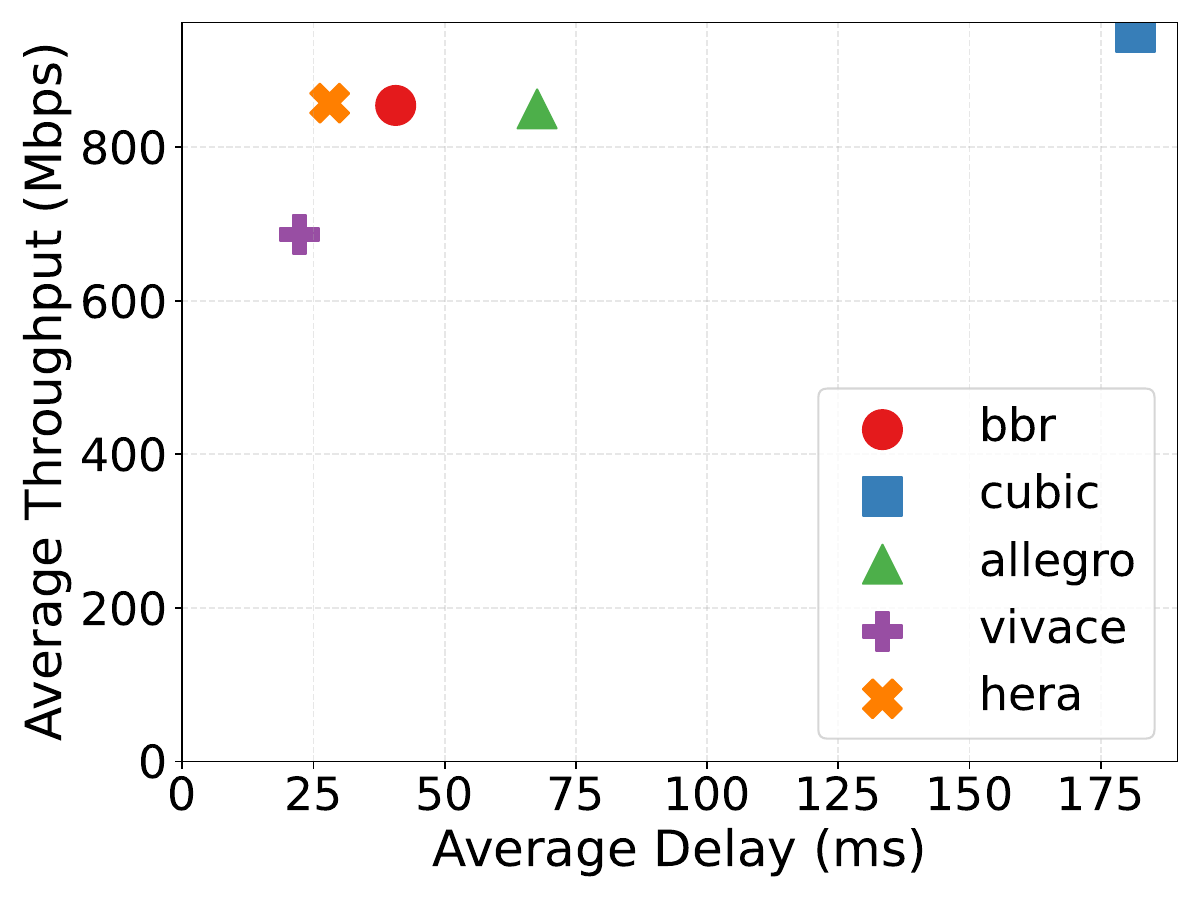}}\label{fig:stationary}\hfill

  \caption{Comparison of congestion control protocols over six 5G network traces, evaluating their average throughput (y-axis) and delay (x-axis). BBR and Cubic achieve high throughput but at the cost of increased latency. PCC Vivace maintains lower delays but struggles under highly fluctuating conditions. PCC Allegro exhibits inconsistent performance, sometimes achieving low delays but often suffering from high latency similar to legacy TCP protocols. \tool\space achieves a balanced performance, maintaining comparable throughput to BBR and Cubic while significantly reducing delay—up to 50\% lower than BBR and 66\% lower than Cubic's delay in some cases. \label{fig:zeusReal3}}
\end{figure*}

\subsection{\tool\space Rate Control Implementation}
\label{sec:\tool-impl}
We implemented the \tool\space  QoE-aware rate control module in the Linux kernel as a pluggable congestion control module~\cite{hemminger2005tcp,pluggable}. The protocol has been tested on Ubuntu 20.04.1 with kernel version 5.15.0-41.
\tool's implementation is open-source.%\footnote{the link will be available upon publication}.

\section{\tool\space  Evaluation}
We aim to compare \tool's performance to other state-of-the-art protocols from several aspects, i.e., the VR streaming performance of the protocol in terms of various VR streaming QoE metrics as well as a traditional throughput-over-delay performance comparison in an emulation environment utilizing the collected 5G traces. 

\subsection{Baseline Selection}
\label{sec:baselines}
We evaluate our congestion control algorithm against four representative baselines that span the spectrum of established and emerging approaches in the context of AR/VR streaming. These include loss-based, model-based, learning-based, and AR/VR-specific schemes, each reflecting different trade-offs between throughput, latency, and stability.

\begin{itemize}[leftmargin=*]
    \item \textbf{Cubic:} The default congestion control algorithm in most operating systems, Cubic represents the class of loss-based controllers. While not optimized for low-latency media, it establishes a widely deployed baseline for throughput and fairness.
    \item \textbf{BBR:} BBR is a modern congestion control algorithm that uses bottleneck bandwidth and minimum RTT estimation to optimize throughput with controlled delay. Its proactive model-based design makes it well-suited for real-time interactive applications.
    \item \textbf{PCC Allegro:} Developed specifically for immersive media streaming, Allegro offers a domain-aware baseline that prioritizes frame delivery deadlines and low buffering, directly aligning with the requirements of AR/VR systems.
    \item \textbf{PCC Vivace:} A reinforcement learning-based algorithm that adapts to dynamic network conditions using latency and throughput feedback. It serves as a strong representative of learned congestion control strategies.
\end{itemize}

We do not include traditional delay-based controllers such as Vegas~\cite{brakmo1995tcp}, LEDBAT~\cite{shalunov2012low}, TIMELY~\cite{mittal2015timely}, or Copa~\cite{arun2018copa}. Vegas and LEDBAT are known to be overly conservative and underperform in shared environments, especially when competing with aggressive loss-based flows like Cubic~\cite{carofiglio2010hands,la1999issues}. TIMELY requires hardware timestamping, which is generally unavailable in wireless and edge-based AR/VR deployments. Copa, while theoretically promising, is not widely adopted and lacks open-source support for immersive media use cases. Additionally, delay awareness is already represented in our chosen baselines—BBR and Vivace both integrate delay feedback using more robust and adaptive mechanisms. Therefore, we focus on practically relevant and competitive baselines tailored to our target domain.

\subsection{Emulated 5G Traces Evaluation\label{sec:zeusres}}
In this subsection, we compare \tool's performance to four other congestion control protocols: Google's BBR, PCC Allegro, PCC Vivace, and the de facto legacy TCP Cubic. We ran each of these protocols across six measured 5G channel traces using MahiMahi. Each protocol was run five times using different seeds to be able to get enough statistical rigor in the results, as well as being able to compute the statistical significance of both the achieved throughput and delay. 

~\Cref{fig:zeusReal3} presents the results comparing \tool\space against the four other chosen congestion control protocols over the emulated 5G traces. 
For the scatter sub-figures, we plot one point per congestion control protocol, corresponding to its measured average throughput (y-axis) and delay (x-axis) combination, averaged over five different seed runs per protocol. A protocol that strongly prioritizes ultra-low latency at the cost of low throughput exists in the bottom-left quadrant of this graph. As the protocol behavior trends towards prioritizing throughput and ``aggressively'' increasing the sending rate, the protocol will gradually move from the bottom-left quadrant towards the top-left quadrant until the protocol starts sending data beyond the available channel capacity, causing bufferbloat-induced queueing delays, which will cause the protocol to move towards the top-right of this graph. In other words, the optimization goal of a protocol determines its place on this curve. For most applications, the ideal operating point on this curve is in the top-left quadrant, only slightly below the area where the curve starts moving towards the top-right, which indicates an increase in latency. This is especially true for \mir applications that demand high bandwidth and also ultra-low latency in order to maintain a smooth and high-quality QoE. Several key takeaways can be seen across the scatter comparison results of the six traces. 

Legacy \ccp tends to have the highest average throughput compared to the rest of the protocols. However, this throughput comes at the expense of much higher delays, often achieving more than three times higher delays compared to other protocols. 

Google's BBR achieves a similar average throughput to TCP Cubic while lowering the delay significantly compared to Cubic. In all experiments, BBR hovers around the ideal top left quadrant. In all of the experiments, Hera is able to achieve lower delays than BBR except for two cases where the delay performance is almost identical. This delay performance improvement of up to 50\% in certain cases comes at a cost of a less than 10\% reduction in throughput performance when compared against BBR.

PCC Vivace generally achieves low delays except in the two most highly fluctuating traces, City Driving and WiGig Walking, which cause Vivace to suffer large delays due to it being unable to adapt to the high fluctuations. In two scenarios, Vivace is able to outperform both Hera and BBR in terms of delay performance for up to a 30\% decrease in throughput utilization. 

PCC Allegro has an inconsistent performance, achieving relatively low throughput and delays in certain traces. In the general case, it tends to achieve high throughput but with high delays as well, similar to the legacy \ccp Cubic in half of the tested scenarios.

Finally, \tool\space achieves a good balance across the different traces, achieving almost the same throughput across the six traces as BBR and Cubic while maintaining lower delays, about half of that of BBR in some cases, such as UMa, RMa, and most notably in the Beach Stationary trace, where \tool\space also manages to outperform other protocols in throughput while still maintaining lower delays. This ability to maintain high throughputs while achieving lower delays enables \tool\space to outperform the other \ccp in terms of VR streaming performance, as we observe in Section~\ref{sec:realvr}.

\subsubsection{Observable QoE Implications}

While the preceding analysis focuses on throughput and latency, these network metrics translate directly into observable Quality of Experience (QoE) characteristics for users in collaborative XR applications. Table~\ref{tab:qoe-metrics} summarizes key QoE metrics that can be inferred from throughput and latency measurements and how they manifest in user experience. In our experiments, the different congestion control protocols display varying throughput-latency tradeoffs that map directly to observable QoE outcomes in collaborative XR applications:

\textbf{Cubic}: While Cubic achieves high average throughput, it does so at the cost of substantial latency increases, often exceeding acceptable limits for interactive MIR applications. As a result, users would experience frequent delays in seeing their collaborators' actions reflected in the shared space, leading to impaired collaborative fluency and responsiveness. The high latency also increases motion-to-photon delays and may contribute to discomfort or disorientation.

\textbf{BBR}: BBR generally balances high throughput with lower latency than Cubic, enabling better QoE. Users are likely to experience smoother interactions and fewer stalls, with video resolution often maintained at higher levels. However, under certain highly variable network conditions, BBR may still cause latency spikes, resulting in momentary interaction lag.

\textbf{PCC Allegro}: Allegro shows inconsistent throughput and latency behavior across traces. In scenarios where it behaves aggressively, it may cause similar QoE degradation to Cubic, such as higher stalls and lower collaborative fluency. In other cases, its lower throughput would force the system to reduce video resolution, leading to a loss of visual fidelity without significant latency gains.

\textbf{PCC Vivace}: Vivace is typically able to maintain lower latency, resulting in good responsiveness and collaborative fluency in stable conditions. However, in highly fluctuating environments (e.g., City Drive, Indoor Walking traces), Vivace struggles to adapt, leading to sharp QoE degradation through either increased buffering or reduced resolution to compensate for throughput drops.

\textbf{Hera}: Hera consistently achieves low latency while maintaining high throughput across diverse network conditions. This directly supports superior QoE: fast startup times, minimal stall events, sustained high video resolution (e.g., 4K where possible), and fluid, responsive collaborative interactions. Hera's design prioritizes latency stability, making it particularly effective at preserving immersion and reducing discomfort in XR environments.

\begin{figure*}[t]
    \centering
    \subfloat[Average Startup Delay (s)]{
        \includegraphics[width=0.3\linewidth]{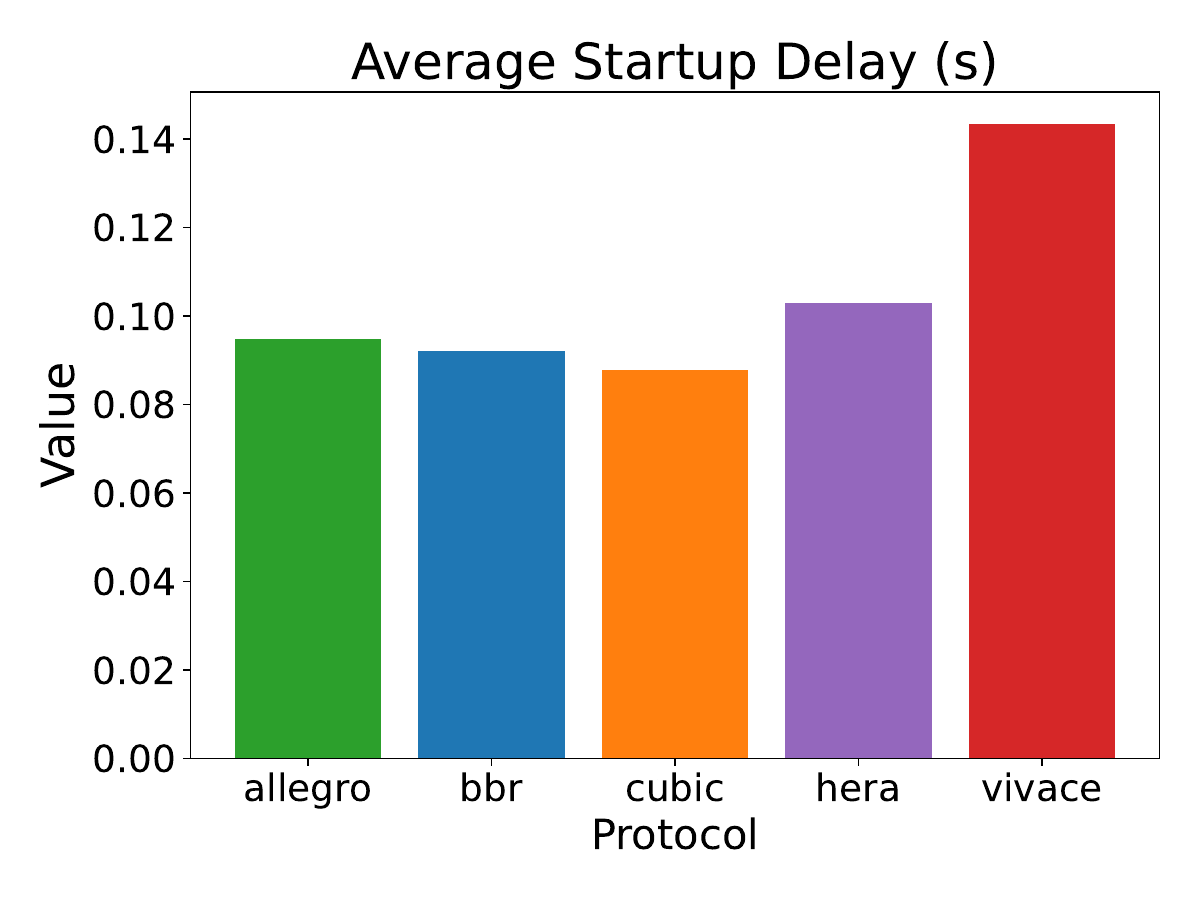}
    }
    \hfill
    \subfloat[Average Collaborative Fluency]{
        \includegraphics[width=0.3\linewidth]{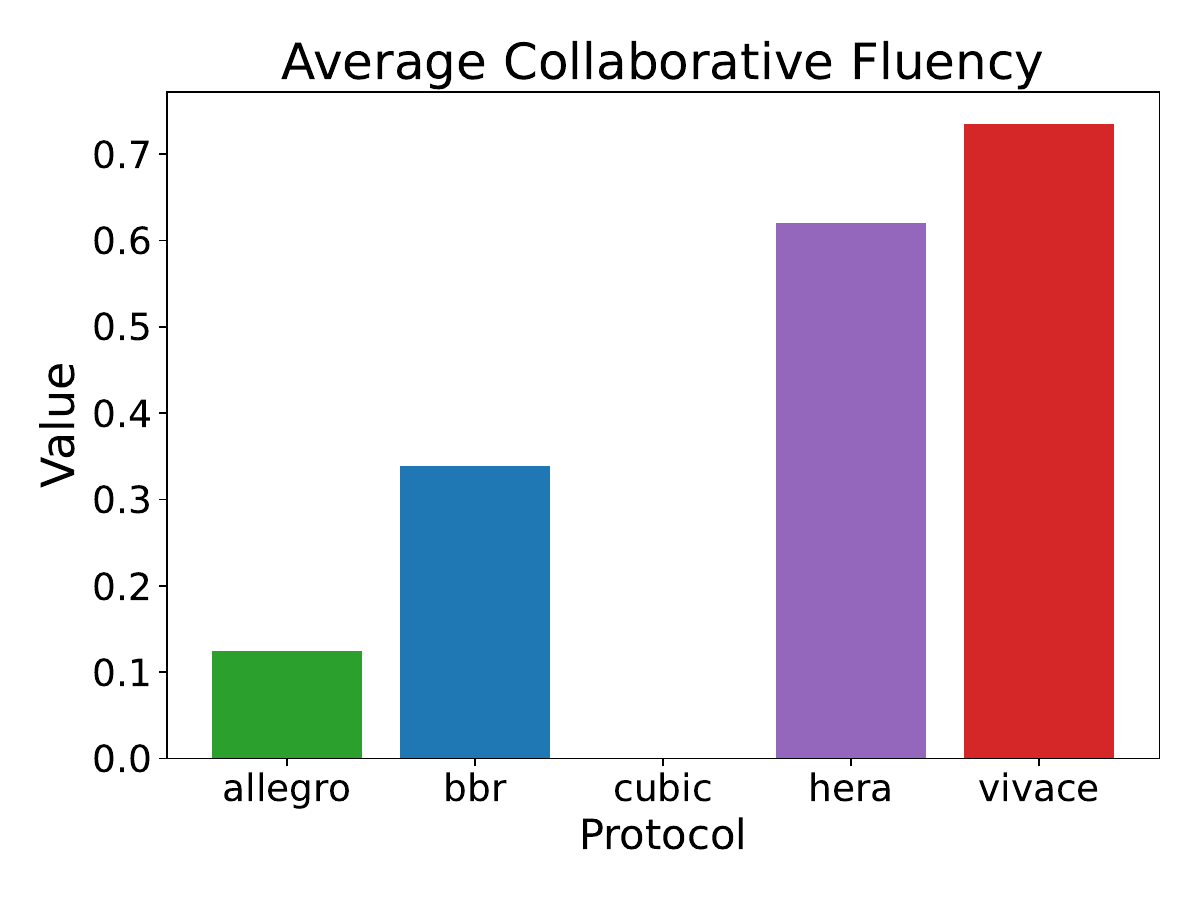}
    }
    \hfill
    \subfloat[Average Interaction Latency (ms)]{
        \includegraphics[width=0.3\linewidth]{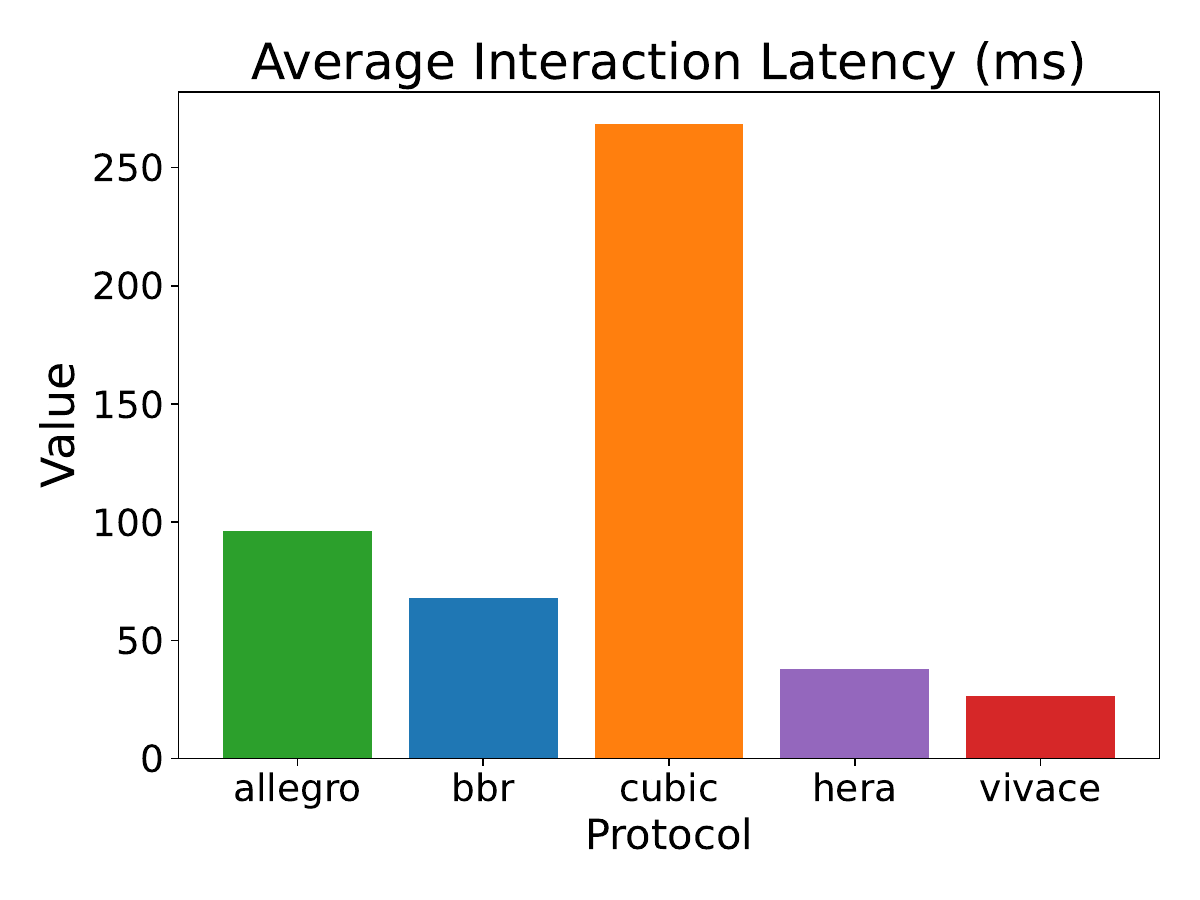}
    }
    \caption{Comparison of QoE metrics across protocols: startup delay, collaborative fluency, and interaction latency.}
    \label{fig:qoe_metrics}
\end{figure*}

To visualize this mapping, Figure~\ref{fig:qoe_metrics} summarizes the relative QoE performance of the tested protocols across key metrics, including startup delay, resolution, and interaction responsiveness, derived from their throughput-latency characteristics in our experiments.

\subsection{VR Multi-User Application\label{sec:realvr}}

\begin{figure}[!hbt]
    \centering
    \subfloat{\includegraphics[width=\linewidth]{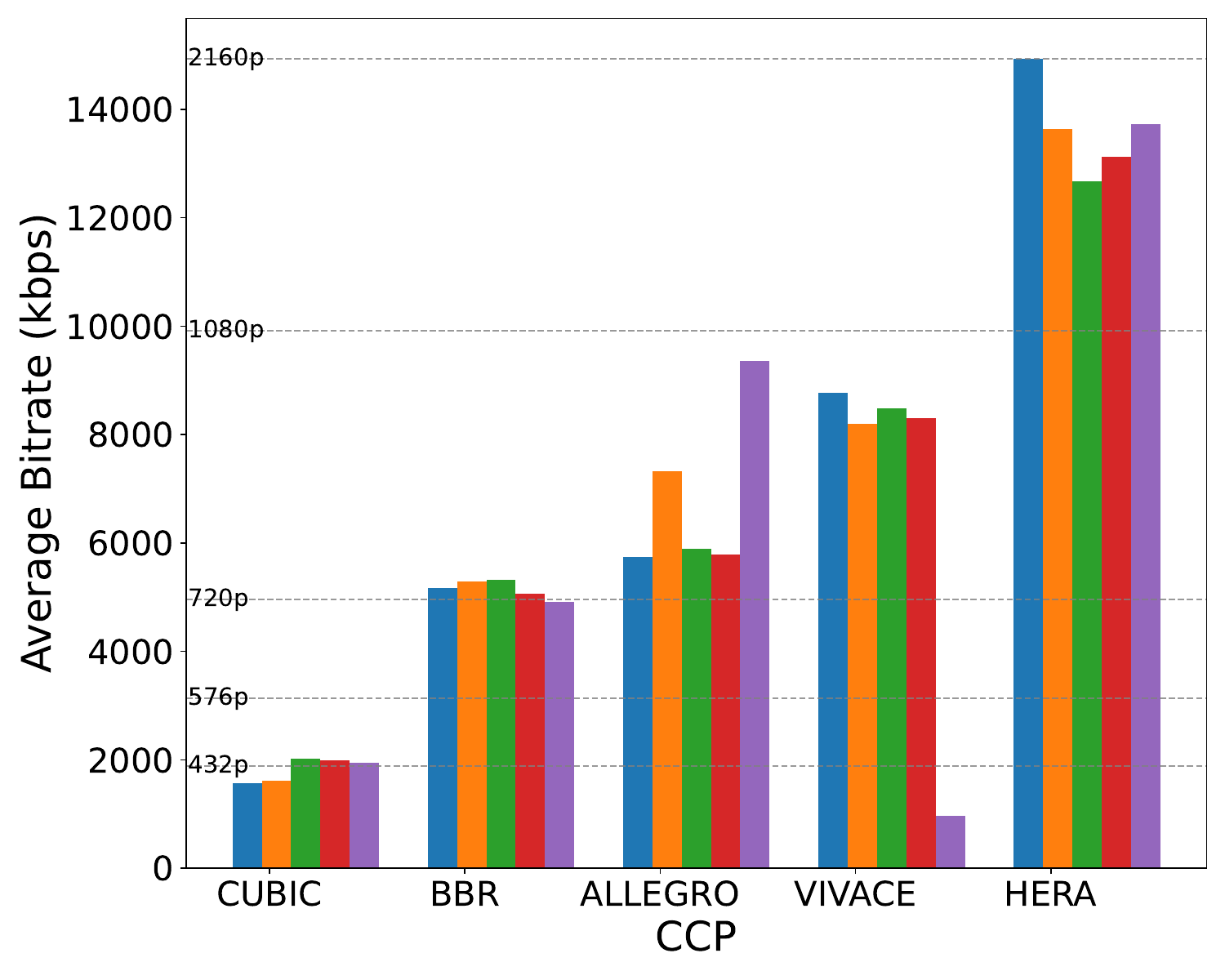}}\hfill
    \caption{Average bitrate of protocols in Mixed Reality streaming applications implemented on Meta Quest headsets. Five colors represent five simultaneous streams. \tool\space maintains 4K quality, while BBR and Cubic degrade to lower resolutions due to congestion. Allegro and Vivace perform better but show fairness issues, leading to inconsistent QoE.}
    \label{fig:streamdash_res}
\end{figure}

We evaluate the performance of five congestion control protocols when supporting a real Mixed Reality application with users running the application on a range of devices, including Meta Quest headsets. Figure~\ref{fig:streamdash_res} shows the average bitrate, which corresponds to the video playback resolution, of five simultaneous streams in a realistic network environment with different congestion control protocols supporting the server. As the videos are encoded in multiple levels of quality, when the system detects that the video buffering rate is low due to congestion, the Adaptive Bitrate Algorithm reduces the video quality. The results show \tool\space outperforming existing protocols and consistently maintaining 4K stream quality for all clients, while BBR drops to 720p stream quality and Cubic performs even worse due to packet losses in the network environment. Allegro and Vivace perform slightly better for some users, but we observe that both protocols have outliers, which indicates fairness issues causing the QoE for different users to vary. The experiments in Figure~\ref{fig:zeusReal3} illustrate that \tool\space consistently achieves lower throughput than some protocols. Despite this, we observe higher streaming QoE for \tool\space due to the fact that higher latencies incurred by aggressive protocols directly degrade streaming quality by introducing delays in the delivery of video and audio data, causing buffering and consequently causing the ABR algorithm to lower the stream quality. Although our analysis also revealed minor frame drops across all protocols, with \tool exhibiting the lowest and Cubic the highest frequency, these differences did not translate to a noticeable impact on the user experience, as the system dynamically adapts the video quality to ensure minimum interruptions at the cost of video resolution.

\subsection{Fairness Analysis}

\begin{figure}[t]
   \subfloat[BBR. Fairness=0.989]{\includegraphics[width=0.49\columnwidth]{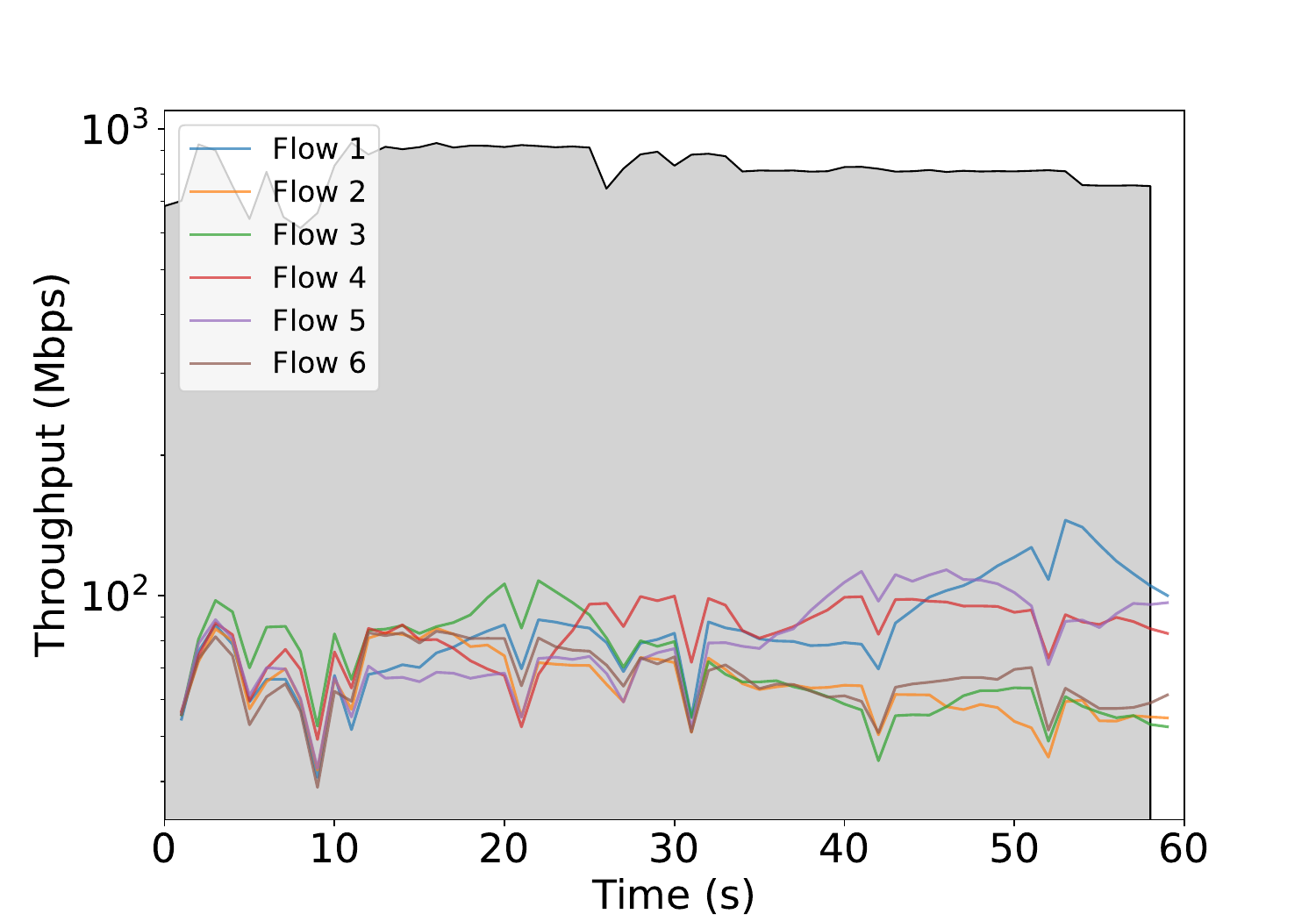}\label{fig:bbrfairness}}
  \subfloat[Allegro. Fairness=0.37]{\includegraphics[width=0.49\columnwidth]{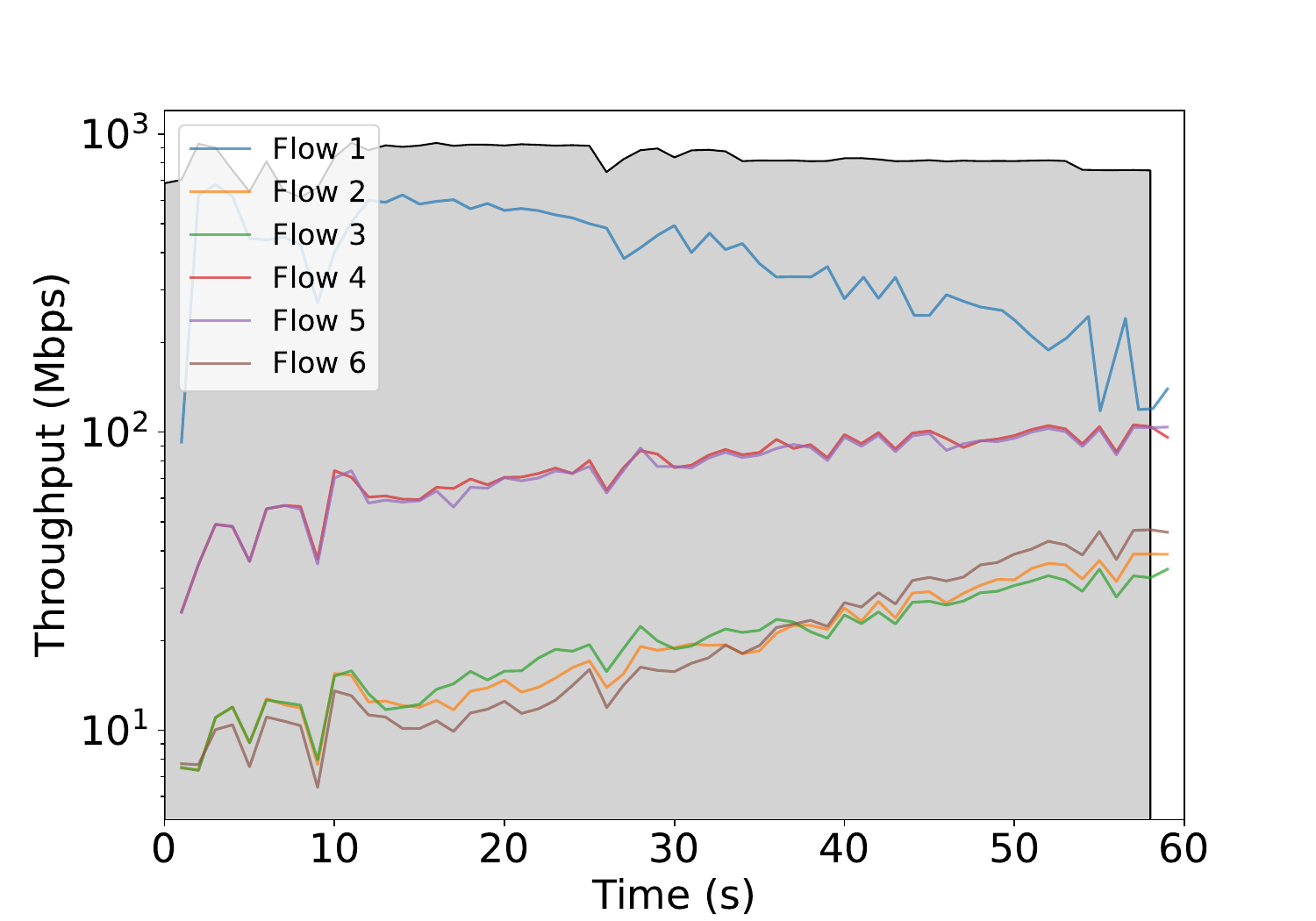}\label{fig:alegrofairness}}\hfill
  \subfloat[Cubic. Fairness=0.999]{\includegraphics[width=0.49\columnwidth]{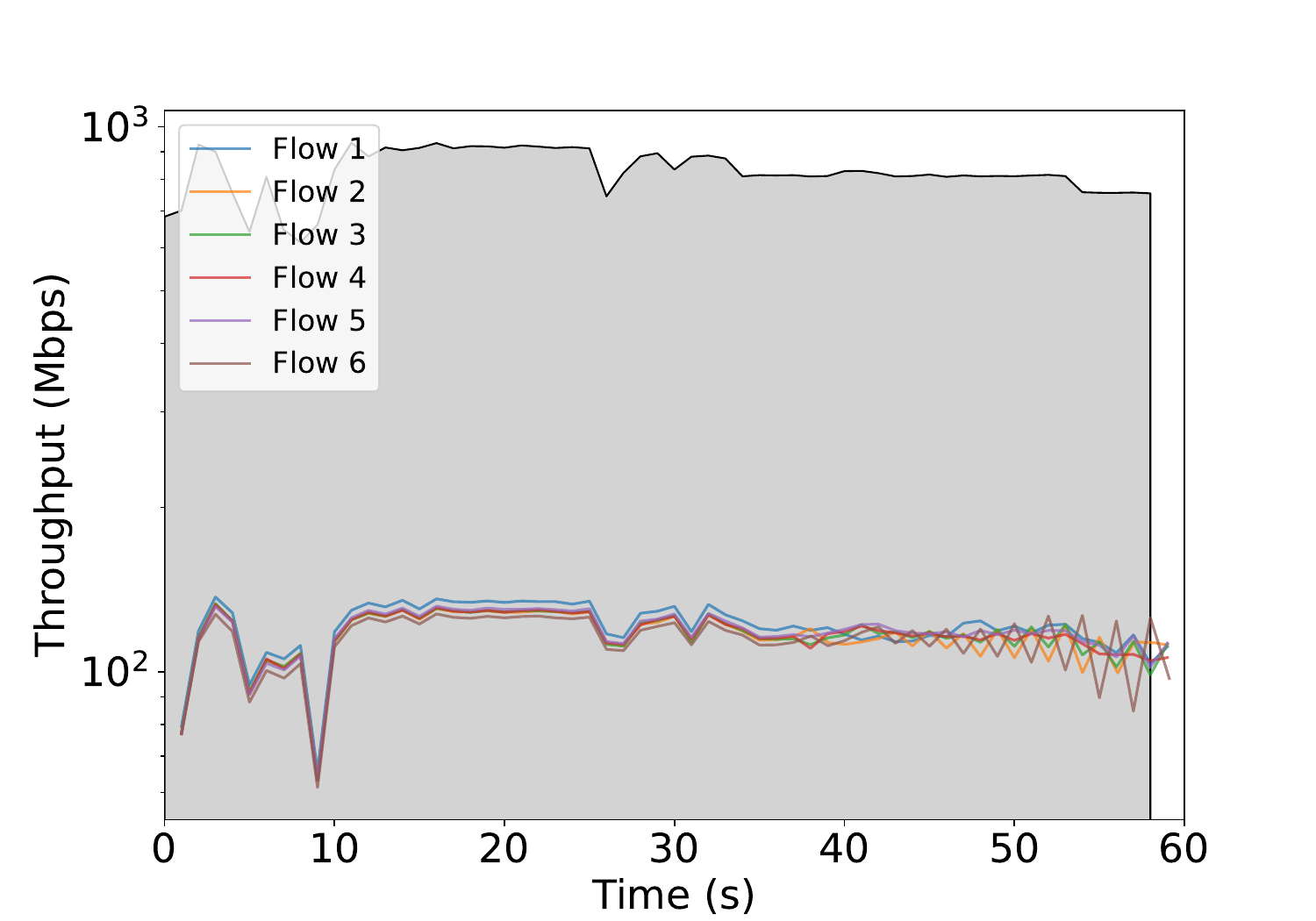}\label{fig:cubicfairness}}
  \subfloat[Vivace. Fairness=0.803]{\includegraphics[width=0.49\columnwidth]{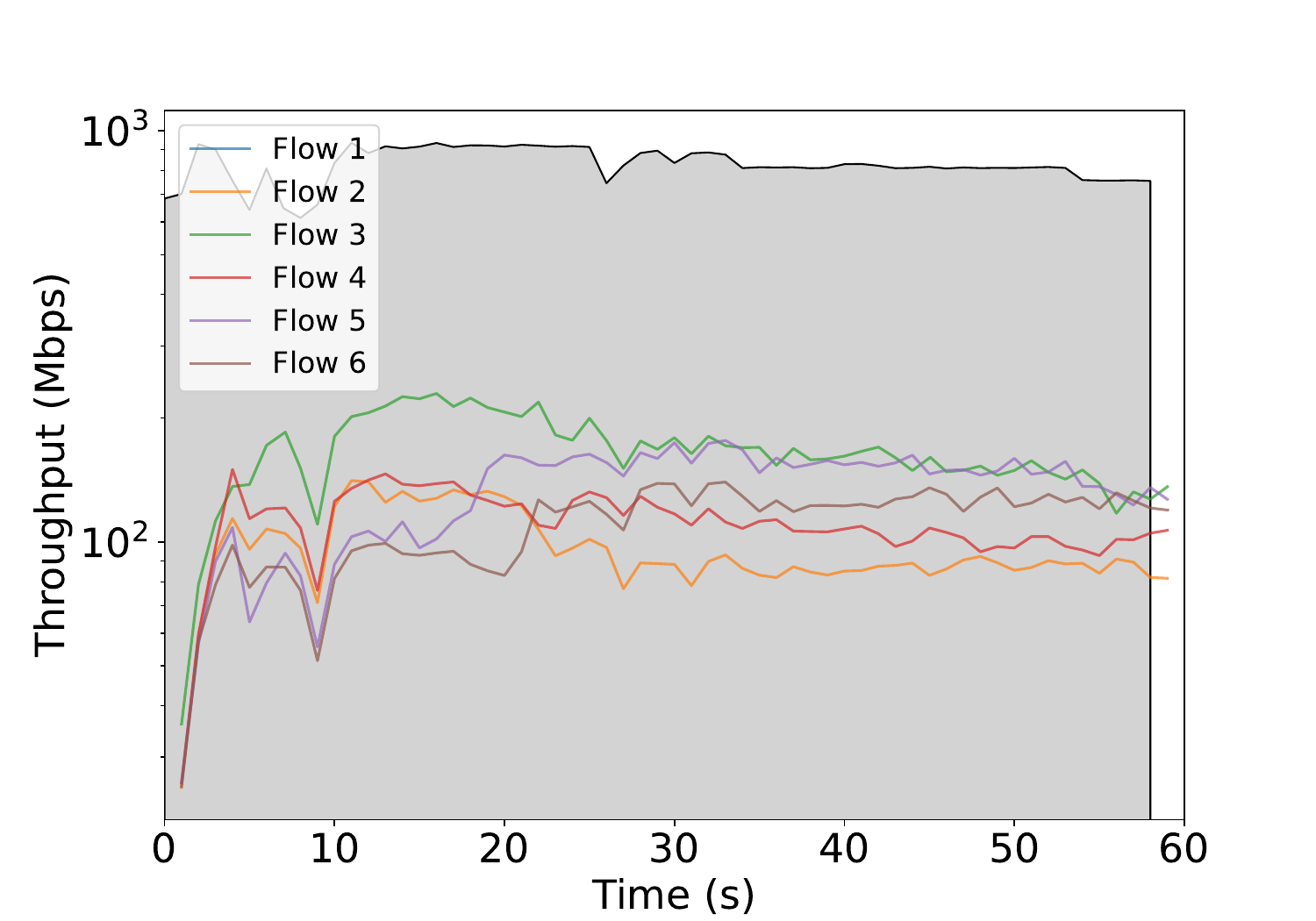}\label{fig:vivacefairness}}\hfill
  \subfloat[Hera. Fairness=0.965]{\includegraphics[width=0.49\columnwidth]{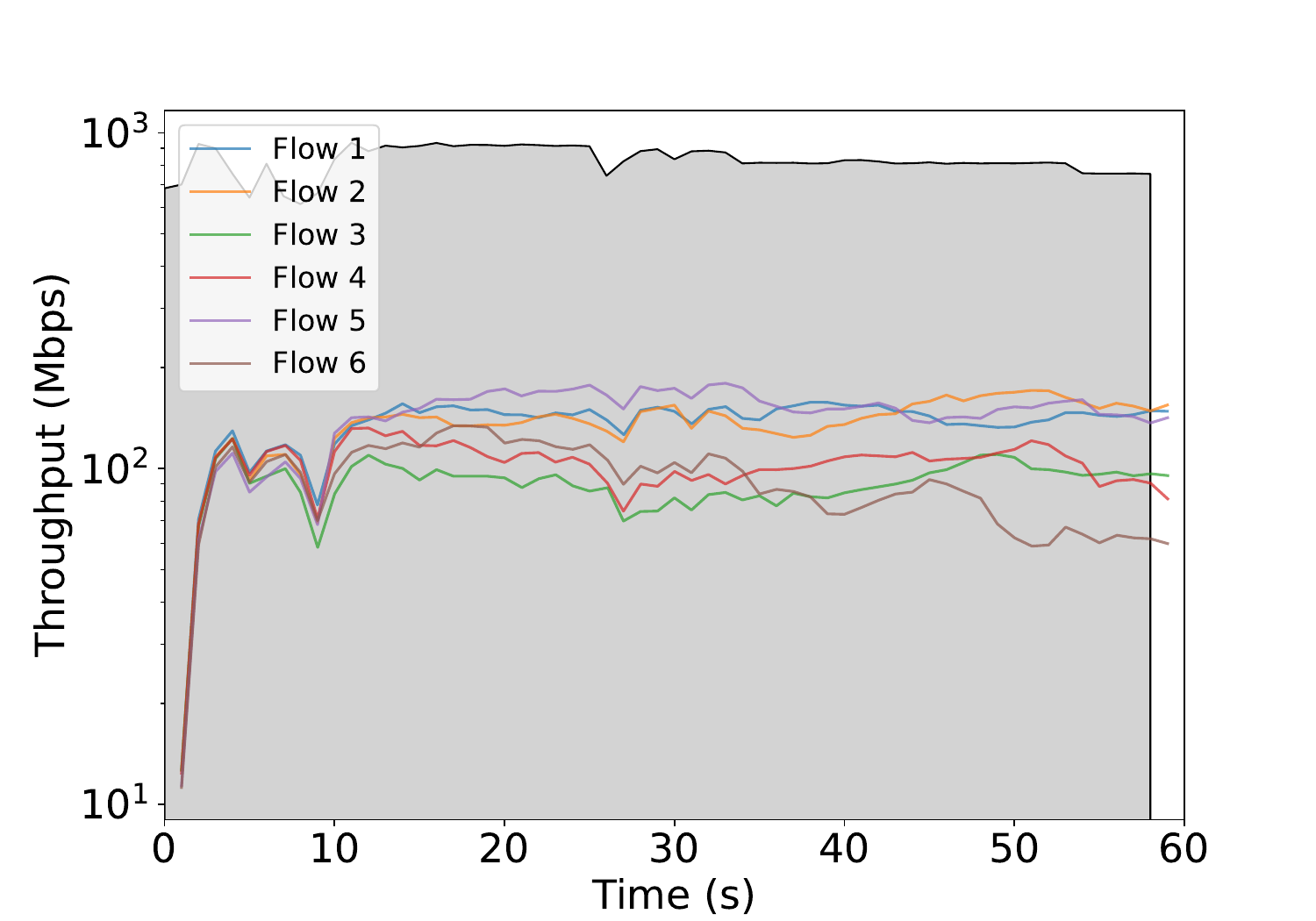}\label{fig:herafairness}}\hfill
  \caption{Throughput distribution across six competing network flows for different congestion control protocols in a 5G network environment. Cubic, BBR, and \tool\space achieve high fairness (Jain's fairness index > 0.96) by evenly distributing available bandwidth. Allegro and Vivace, however, show poor fairness, with Allegro exhibiting extreme disparities where certain flows dominate bandwidth, and Vivace struggling with significantly lower throughput across flows.\label{fig:fairness}}
\end{figure}

% BBR Jain Fairness Index: 0.988785545814478
% allegroJain Fairness Index: 0.36955053886113387
% vivaceJain Fairness Index: 0.8029299678617425
% cubicJain Fairness Index: 0.9997313264813299
% heraJain Fairness Index: 0.9645282355483105

A critical component of any congestion control protocol is the ability to operate in an environment with multiple network flows using different \ccp. This is especially true in \mir applications where multiple users simultaneously use the application and compete for network resources. As such, we evaluate the Jain fairness index~\cite{jain1984quantitative} for all \ccp for traditional as well as \mir applications. As discussed in Section~\ref{sec:realvr} where we observe the performance of different protocols supporting five competing clients accessing the application content simultaneously. We observe that Cubic and BBR display strong intra-fairness, while PCC Allegro and Vivace show outliers that lower the fairness score. \tool\space demonstrates slightly lower fairness than Cubic and BBR but with significantly improved video quality for each user. 

Additionally, we conduct various experiments using our experimental framework to analyze the intra-fairness behavior of the selected \ccp when operating in a 5G network environment to replicate the scenario where multiple users in a local environment are simultaneously accessing the application and competing for network resources in a 5G network environment. For traditional scenarios, as discussed in Section~\ref{sec:zeusres}, we evaluate the behavior of each \ccp when operating in the \textit{Beach Stationary} simulated 5G network environment using the network trace collected from a real 5G deployment in the wild. The results of these experiments are illustrated in Figure~\ref{fig:fairness}. For our experiments, we run 6 competing flows and observe the fairness characteristics of our protocol in terms of throughput allocated over time to each flow. We also employ the widely used Jain's fairness index~\cite{jain1984quantitative} metric to quantify the overall fairness characteristics of each \ccp in these experiments. The results corroborate the findings from the \mir streaming experiments. In our experiments, Cubic, BBR, and Hera demonstrate strong fairness characteristics with above 90\% Jain's fairness, allocating an equal share of the available bandwidth across all connected clients, with Allegro and Vivace falling behind due to a small subset of the flows dominating the other flows in the case of Allegro, or performing significantly worse in terms of throughput compared to the other flows in Vivace.

\section{Discussion}
\label{sec:discussion}

\subsubsection*{Performance Discussion}
Traditional congestion control algorithms exhibit several fundamental weaknesses when applied to MIR streaming over 5G networks, leading to suboptimal performance compared to Hera. Conventional loss-based TCP algorithms such as Cubic and Allegro assume that packet loss signals network congestion. However, in cellular networks, packet losses frequently occur due to handoffs, signal fluctuations, and interference, rather than genuine congestion. This results in unnecessary rate reductions, degrading throughput and streaming quality. Also, the end-to-end nature of TCP means that congestion response happens far from the point of signal fluctuation, leading to delayed adaptation~\cite{mangiante2018congestion}. In highly dynamic 5G environments, where channel conditions fluctuate on millisecond timescales, algorithms like BBR struggle to converge because they rely on multi-RTT bandwidth estimation windows (6-10 RTTs)~\cite{cardwell2016bbr}. This results in slow reactions to rapid network changes, leading to both underutilization and excessive queuing.

Hera overcomes these limitations by employing a histogram-based RTT tracking mechanism to dynamically adjust its congestion window (cwnd), allowing it to proactively respond to latency fluctuations rather than relying on delayed congestion signals. Unlike loss-based algorithms, Hera avoids unnecessary rate reductions due to non-congestive packet losses, which are common in 5G environments. By prioritizing low-latency operation while maintaining stable throughput, Hera prevents both over-congestion, as seen in BBR, and under-utilization, which affects protocols like Allegro and Vivace in highly variable networks. These optimizations allow Hera to deliver consistent high video quality with minimal delay, making it particularly effective for bandwidth-intensive, latency-sensitive MIR applications.

\section{Conclusion}
\label{sec:conclusion}

This paper presents \tool, a modular framework that integrates a novel QoE-aware rate control protocol with a high-level AR/VR application layer designed to meet the demanding requirements of multi-user immersive reality experiences over next-generation wireless networks. Through extensive evaluation using realistic 5G network scenarios and a custom-built streaming and benchmarking system, we demonstrate that \tool\space consistently achieves lower latency, higher throughput, and greater fairness compared to state-of-the-art congestion control protocols including BBR, Allegro, Vivace, and Cubic. By combining the low-level congestion control module with an application-aware streaming and synchronization layer, \tool\space enables dynamic adaptation of video resolution, frame rate, and collaborative update rates in response to real-time network conditions. Our results show that this integrated design significantly improves key Quality of Experience (QoE) metrics such as startup delay, stall frequency, and responsiveness that are critical to sustaining high-quality multi-user AR/VR sessions. Overall, this work illustrates the requirements to unlock the full potential of immersive multi-user experiences on emerging 5G and future wireless infrastructures. Future work will explore scaling the framework to larger user groups, incorporating edge computing support, and extending the congestion control techniques to accommodate new transport protocols and network architectures.

\subsubsection*{Ethics Statement} This work does not raise any ethical issues.

\balance
\bibliographystyle{ACM-Reference-Format}
\bibliography{references}


\end{document}